\documentclass[prb,
 superscriptaddress, twocolumn,
aps,showpacs]{revtex4-1} 

\usepackage{graphicx}
\usepackage{hyperref}
\usepackage{amsmath}
\usepackage{url}
\usepackage{color}
\usepackage{verbatim}
\usepackage{gensymb}

\def\beq {\begin{equation}}
\def\eeq {\end{equation}}
\def\w {\omega}
\def\bfq {\mathbf{q}}

\def\bfG {\mathbf{G}}
\def\bfk {\mathbf{k}}
\def\bfr {\mathbf{r}}

\date{\today}
\begin{document}

\title{Excitons in van der Waals materials: from monolayer to bulk hexagonal boron nitride}

\newcommand{\lsi}{Laboratoire des Solides Irradi\'es, \'Ecole Polytechnique, CNRS, CEA,  Universit\'e Paris-Saclay, F-91128 Palaiseau, France}
\newcommand{\etsf}{European Theoretical Spectroscopy Facility (ETSF)}
\newcommand{\soleil}{Synchrotron SOLEIL, L'Orme des Merisiers, Saint-Aubin, BP 48, F-91192 Gif-sur-Yvette, France}
\newcommand{\helsinki}{Department of Physics, P.O. Box 64, FI-00014 University of Helsinki, Helsinki, Finland}

\author{Jaakko Koskelo}
\affiliation{\helsinki}

\author{Giorgia Fugallo}
\affiliation{\lsi}
\affiliation{\etsf}

\author{Mikko Hakala}
\affiliation{\helsinki}

\author{Matteo Gatti}
\affiliation{\lsi}
\affiliation{\etsf}
\affiliation{\soleil}

\author{Francesco Sottile}
\affiliation{\lsi}
\affiliation{\etsf}

 \author{Pierluigi Cudazzo}
 \affiliation{\lsi}
\affiliation{\etsf}
 
\begin{abstract}
We present a general picture of the exciton properties of layered materials 
in terms of the excitations of their single-layer building blocks.
To this end, we derive a model excitonic hamiltonian by drawing an analogy 
with molecular crystals, which are other prototypical van der Waals materials.
We employ this simplified model to analyse in detail the excitation spectrum  
of hexagonal boron nitride (hBN) that we have obtained from the {\it ab initio} solution 
of the many-body Bethe-Salpeter equation as a function of momentum. 
In this way we identify the character of the lowest-energy excitons in hBN, 
discuss the effects of the interlayer hopping and the electron-hole exchange interaction on the exciton dispersion,
and illustrate the relation between exciton and plasmon excitations in layered materials.
\end{abstract} 

\pacs{71.35.-y,78.67.-n,78.20.Bh }

\maketitle

\section{Introduction}

In many nanostructured materials, while strong covalent bonding 
provides the stability of the sub-nanometric  elementary units,
the whole assembly is held together by weak van der Waals interactions.
The individual building blocks hence 
maintain most of their intrinsic characteristics
also when arranged together to form a crystalline solid.
In principle, novel materials properties can be thus  tailored  
by controlling those of the elementary units. 
This bottom-up strategy in the synthesis of new materials
has been intensively followed  since the 1980s, 
when small atomic aggregates, nanoclusters,
fullerenes, nanotubes, etc. started to
attract  enormous 
attention \cite{Dresselhaus1996,Dresselhaus2003,Moriarty2001,Klimov2003}.
After the isolation of graphene in the mid 2000s, 
the focus of interest in nanotechnology applications
has largely shifted towards 
two-dimensional (2D) materials \cite{Novoselov2005}.
In recent years, monolayers or few-layer crystals of hexagonal boron nitride (hBN),
black phosphorus,  
transition-metal dichalcogenides,  and several other materials,
have been also heavily investigated \cite{Sheneve2013,Bhimanapati2015}.
The technological challenge now resides in the ability to stack together different atomically thin layers 
in order to build new kinds of ``van der Waals heterostructures'', with the goal 
of realising devices with customized functionalities \cite{Geim2013,Novoselov2016}.

In order to design materials with desired features
for improved  nanoelectronics and optoelectronics applications\cite{Wang2012,Xia2014},
the  
optical properties of layered materials 
need to be understood in detail. 
Due to the reduced effective  screening \cite{Cudazzo2011},
the optical response of 2D materials is dominated by strong electron-hole (e-h) interactions 
giving rise to bound e-h pairs, i.e. excitons.
 Nowadays, the state-of-the-art method to describe excitonic effects in condensed matter  
 is the solution of the Bethe-Salpeter equation (BSE) \cite{Hanke1979,Strinati1988}
 within the GW approximation (GWA) \cite{Hedin1965}  of many-body perturbation theory \cite{Martin2016}.
 As a matter of fact, in the last couple of decades the {\it ab initio} BSE  scheme \cite{Onida1995,Albrecht1998,Benedict1998,Rohlfing1998}
 has been successfully applied to a wide variety of materials, 
 including systems with reduced dimensionality \cite{Rohlfing2000,Onida2002,Martin2016}.

Here, on the basis of {\it ab initio} GW-BSE calculations, 
 we derive a general formalism to describe excitons in 
layered crystals starting from the knowledge of the excitations of a single layer.
To this end, we proceed by analogy with molecular crystals \cite{Cudazzo2012,Cudazzo2013,Cudazzo2015}, 
which can indeed be considered as the prototypical case of van der Waals materials.
In this way we obtain a general picture of excitonic effects in layered systems
in terms of the interplay between 
e-h exchange interaction and band dispersion (i.e. interlayer hopping),
which allows us to distinguish in a simple manner 
excitons of different character 
(e.g. intralayer and interlayer excitons).
To  numerically illustrate our analysis, 
we have chosen a prototypical layered material, namely hexagonal boron nitride, 
for which GW-BSE calculations 
are well established \cite{Blase1995,Cappellini2001,Arnaud2006,Wirtz2006,Arnaud2008,Wirtz2008,Marini2008,Galambosi2011,Fugallo2015,Cudazzo2016,Henck}.
In hBN the calculated dielectric function has already shown to be in excellent agreement
with experiment in a wide range of energy and momentum \cite{Fugallo2015}.
Here we obtain the eigenvalue spectrum of the excitonic hamiltonian as a function of momentum \cite{Gatti2013}
and discuss its relation with quantities that are accessible via experiments.

The present work also extends to the exciton case (via the BSE formalism)
the previous {\it ab initio} investigations
that studied plasmons
(i.e. collective electronic excitations)
in prototypical layered systems like
graphite \cite{Marinopoulos2002,Marinopoulos2004,Hambachphd} or
multilayer graphene \cite{Hambachphd,Wachsmuth2014}.
In those materials, dielectric properties as a function of  momentum $\bfq$ 
and interlayer distance $d$ were calculated in the random-phase approximation (RPA)
within the framework of time-dependent density-functional theory.
Those studies already addressed general questions like
the effects of crystal local fields 
due to spatial inhomogeneities in the charge-density 
variation of the Hartree potential, 
the role of the interlayer coupling due to the
long-range Coulomb interaction between charge oscillations on different layers, 
and the possibility to adopt a local-response approximation to formally 
relate 2D and 3D response functions \cite{Hambachphd}.
More recently, a ``quantum-electrostatic heterostructure'' model \cite{Andersen2015}
was similarly derived to describe the dielectric properties of complex multilayers
starting from those of the single-layer building blocks, also
taking into account the long-range coupling between layers due to the Coulomb interaction. 
However, in both cases hybridisation effects were neglected: in the present work 
they will be analysed in detail in terms of interlayer hopping mechanisms.

\section{Theoretical framework and computational details}
\label{sec:theory}

The BSE is a formally exact Dyson-like equation relating
the electron-hole correlation function $L$ to its independent-particle version $L_0$ \footnote{For an extended introduction to the theoretical background see e.g. Refs. \protect\onlinecite{Martin2016,Onida2002}.}.
Within the GWA to the self-energy, the BSE reads:
\begin{multline} \label{dyson}
 L(1,2,3,4) = L_0 (1,2,3,4) +   L_0(1,2,5,6) \\ \times [v(5,7)\delta(5,6)\delta(7,8) 
 -W(5,6)\delta(5,7)\delta(6,8)]L(7,8,3,4)
\end{multline}
where we have used the shorthand notation $(1)$ for position, time and spin $(\mathbf{r}_1, t_1, \sigma_1)$ and
repeated indices are integrated over. 
In \eqref{dyson} $v$ is the bare Coulomb interaction and $W$ its statically screened version calculated at the RPA level.
The former enters the kernel of the BSE \eqref{dyson} as an e-h exchange repulsive interaction and is responsible for crystal local-field effects.
The latter is a direct attractive e-h interaction that is at the origin of excitonic effects, including the formation of bound excitons.
For triplet excitons the e-h exchange interaction $v$ is absent.

The diagonal of the correlation function $L$ yields the density-density response function $\chi(1,2)=L(1,1,2,2)$.
In a crystal, by taking the Fourier transform of $\chi$ to frequency and reciprocal-lattice space, 
one directly obtains the loss function $-\rm{Im} \epsilon_M^{-1}$ as:
\beq
-\rm{Im} \epsilon_M^{-1}(\bfq,\omega) = -\frac{4\pi}{q^2} \rm{Im} \chi(\bfq,\bfq,\w).
\eeq
Here $\epsilon_M$ is the macroscopic dielectric function and $\bfq$ is a wave vector such that $\bfq=\bfq_r+\bfG_0$,
where $\bfq_r$ belongs to first Brillouin zone (1BZ) and $\bfG_0$ is a reciprocal-lattice vector.
The loss function, which can be measured by inelastic x-ray scattering (IXS) and electron energy loss 
spectroscopy (EELS), 
describes the longitudinal linear response of the system to an external potential. It gives
hence  access  to  collective  excitations  such  as  plasmons,  and (screened) electron-hole excitations.

Optical absorption spectra are related to the vanishing-$\bfq$ limit of $\rm{Im} \epsilon_M(\bfq,\omega)$, 
which can be obtained from the Fourier transform of the modified response function $\bar \chi(1,2)=\bar L(1,1,2,2)$:
\beq
\rm{Im} \epsilon_M(\bfq,\omega) = -\frac{4\pi}{q^2} \rm{Im} \bar \chi(\bfq,\bfq,\w),
\eeq
where $\bar L$ satisfies the BSE \eqref{dyson} with the modified Coulomb interaction $\bar v$ at the place of $v$.
In the reciprocal space, $\bar v$ is defined to be equal to $v$ except for the $\bfG_0$ component for which it is set to 0\cite{Gatti2013}.
Therefore, the difference between optical absorption  and loss function at $\bfq=0$ is given by the long-range $\bfG_0=0$ 
component of the Coulomb interaction\cite{Onida2002,Sottile2005,Sottilephd}
[which is absent for  $\rm{Im} \epsilon_M$ in the BSE \eqref{dyson}].

The loss function can be also explicitly written in terms of the imaginary and real parts of the macroscopic dielectric function:
\begin{equation} \label{lossf}
-\rm{Im} \epsilon_M^{-1}(\bfq,\omega)=\frac{\rm{Im} \epsilon_M(\bfq, \omega)}{[\rm{Re} \epsilon_M(\bfq, \omega)]^{2}+[\rm{Im} \epsilon_M(\bfq, \omega)]^2}.
\end{equation}
Plasmon excitations are peaks in $-\rm{Im} \epsilon_M^{-1}$ corresponding to the frequencies where $\rm{Re} \epsilon_M$ is zero and 
$\rm{Im} \epsilon_M$ (which provides the damping of the plasmon) is not too large.

In order to describe correlated e-h pairs explicitly, the BSE \eqref{dyson} (with $\bar v$ at the place of $v$)
can be cast in the form of a two-particle Schr\"odinger equation with an excitonic hamiltonian: 
\begin{multline} \label{eqhex}
\hat{H}_{ex}(\bfq) = \sum_{c \bfk}E_{c \bfk+\bfq} a^{\dag}_{c\bfk+\bfq} a_{c\bfk+\bfq}-\sum_{v\bfk}E_{v\bfk} b^{\dag}_{v\bfk} b_{v\bfk} \\
+ \sum_{\substack{vc\bfk, \\ v'c'\bfk'}} 
\left[ 2 \delta_m \bar v_{v'c'\bfk'}^{vc\bfk}(\bfq) - W_{v'c'\bfk'}^{vc\bfk}(\bfq)  \right]
a^{\dag}_{c\bfk+\bfq}b^{\dag}_{v\bfk}b_{v'\bfk'}a_{c'\bfk'+\bfq}.
\end{multline}
Here $\bfk$, belonging to the 1BZ, and $v$($c$) denote a valence (conduction) Bloch state of energy $E_{v\bfk}$ ($E_{c\bfk}$) calculated within the GWA;
$a^{\dag}$ ($a$) and  $b^{\dag}$ ($b$) are creation
(annihiliation) operators for electrons and holes, respectively;
$\delta_m$ is 1 for the singlet and 0 for the triplet channel.

The first line of \eqref{eqhex} is an independent particle hamiltonian $\hat H_{ip}$ (corresponding to $L_0$ in the Dyson equation \eqref{dyson}), 
while the second line contains the interaction terms stemming from the kernel of \eqref{dyson}. 
The matrix elements of $\bar v$ and $W$ are calculated in the basis of Bloch states  as\cite{Rohlfing2000,Gatti2013}:
\begin{align}
 \bar v_{v'c'\bfk'}^{vc\bfk}(\bfq) = &
 \int d\mathbf{r}d\mathbf{r}^\prime
\psi^*_{c\mathbf{k}+\bfq}(\mathbf{r})\psi_{v\mathbf{k}}(\mathbf{r}) \bar v(\mathbf{r},\mathbf{r}^\prime) \nonumber \\
& \times \psi_{c^\prime\mathbf{k}^\prime+\bfq}(\mathbf{r}^\prime)\psi^*_{v^\prime\mathbf{k}^\prime}(\mathbf{r}^\prime)
\end{align}
\begin{align}
W_{v'c'\bfk'}^{vc\bfk} (\bfq) = &
\int d\mathbf{r}d\mathbf{r}^\prime
\psi^*_{c\mathbf{k}+\bfq}(\mathbf{r})\psi_{c^\prime\mathbf{k}^\prime+\bfq}(\mathbf{r})W(\mathbf{r},\mathbf{r}^\prime) \nonumber \\
& \times \psi_{v\mathbf{k}}(\mathbf{r}^\prime)\psi^*_{v^\prime\mathbf{k}^\prime}(\mathbf{r}^\prime).
\end{align}
In Eq. \eqref{eqhex} we have adopted the Tamm-Dancoff approximation (TDA), 
which amounts to neglecting antiresonant $c\rightarrow v$ transitions and their coupling with resonant $v \rightarrow c$ transitions
(extension to the general case can be seen in \cite{Gatti2013}).

The macroscopic dielectric function 
\begin{equation} \label{epm}
\epsilon_M ({\bf q}, \omega)  =1- \frac{8 \pi}{q^2} \sum\limits_{\lambda} 
 \frac{\left| \sum\limits_{vc{\bf k}} A^{\lambda}_{vc{\bf k}} ({\bf q}) \tilde \rho_{vc\bfk}(\bfq) \right|^2}{\omega - E^{\lambda}(\bfq) + i\eta},
\end{equation}
with the oscillator strengths $\tilde\rho_{vc\bfk}(\bfq)$ defined as:
\beq
\tilde\rho_{vc\bfk}(\bfq) = \langle v{\bf k}-{\bf q}_r  |   \rm{e}^{-i {\bf q} \cdot {\bf r}} | c{\bf k} \rangle,
\eeq
and the exciton wavefunction
\beq
|\Psi^{\lambda} (\bfq)\rangle = \sum_{vc\bfk} A_{\lambda}^{vc\bfk}a^{\dag}_{c\bfk} b^{\dag}_{v\bfk+\bfq} |0 \rangle,
\eeq
where $\bfq$ is the total momentum of the two-particle state,
can be thus written in terms of the eigenvectors $A^{\lambda}(\bfq)$ and the eigenvalues  $E^\lambda(\bfq)$ of the excitonic hamiltonian $\hat{H}_{ex}$ \eqref{eqhex}:
\beq
\hat{H}_{ex}(\bfq)A^{\lambda}(\bfq)=E^\lambda(\bfq)A^{\lambda}(\bfq).
\eeq
The excitonic eigenvalues $E^\lambda(\bfq)$ of $\hat{H}_{ex}$ are hence the poles of the $\bar L$ and $\epsilon_M$ functions in the frequency domain.
They give rise to peaks in the spectrum of $\rm{Im} \epsilon_M(\bfq,\omega)$ whose intensity is given by the numerator of Eq. \eqref{epm}.
If it is zero, the corresponding excitonic state is said to be dark.

The inverse macroscopic dielectric function $\epsilon_M^{-1}$ can be analogously obtained from the eigenvectors and eigenvalues of 
the excitonic hamiltonian $H_{ex}'$ that, in addition to $H_{ex}$ \eqref{eqhex}, also includes the long-range component of the Coulomb interaction:
\begin{equation}\label{epm-1}
\epsilon_M^{-1} ({\bf q}, \omega)  =1 + \frac{8 \pi}{q^2} \sum\limits_{\lambda} 
 \frac{\left| \sum\limits_{vc{\bf k}} A^{'\lambda}_{vc{\bf k}} ({\bf q}) \tilde \rho_{vc\bfk}(\bfq) \right|^2}{\omega - E^{'\lambda}(\bfq) + i\eta}.
\end{equation}
Therefore also the loss function $-\rm{Im} \epsilon_M^{-1} ({\bf q}, \omega)$ can be decomposed in terms of the eigenvalues $E^{'\lambda}(\bfq)$ and the eigenvectors $A^{'\lambda}(\bfq)$ of $H_{ex}'$.

In our first-principles calculations we obtain the single-particle states $\psi_{n\mathbf{k}}$ using Kohn-Sham (KS) density-functional theory within the local-density approximation (LDA) \cite{Kohn1965}. 
We use Troullier-Martins pseudopotentials \cite{Troullier1991}, and expand the KS wavefunctions in a plane-wave basis set with a
cutoff of 30 Hartree. 
The lattice parameters for the bulk  are optimized using the LDA. 
We also consider hBN systems with variable interlayer distances $d$ for which 
the in-plane lattice vectors are kept constant to the bulk value. 
On the basis of the results of GW calculations for hBN bulk \cite{Henck},
we apply a scissor operator of 1.96 eV to correct for the LDA underestimation of the single-particle band gap.
For larger interlayer distances $d$ the GW correction increases\cite{Wirtz2006}: for example it becomes 2.47 eV for $d=1.5 d_0$, where $d_0$ is the interlayer separation of the bulk.
For the GW-BSE computational details of the hBN monolayer we refer to Ref. \onlinecite{Cudazzo2016}.
In all the other cases, we sample the Brillouin zone using a 48$\times$48$\times$4 $\Gamma$-centered grid. 
For the BSE calculations at finite $\bfq$ we follow the same procedure as described in Ref. \onlinecite{Fugallo2015}.
To simplify the analysis of the results in Sec. \ref{sec:discussion}, here we use a minimal e-h transition basis set
comprising 2 valence and 2 conduction bands and solve the BSE within TDA.  
As a consequence of the Kramers-Kronig relations, $\rm{Re} \epsilon_M$  converges more slowly than $\rm{Im} \epsilon_M$
with the number of higher-energy e-h transitions and  (especially at small $\bfq$) is affected 
by the coupling with antiresonant transitions \cite{Olevano2001} neglected in the TDA.
While in the present case the main interest is to establish a direct connection between the electronic excitations characterising
$\rm{Im} \epsilon_M ({\bf q}, \omega)$ 
and the loss function $-\rm{Im} \epsilon_M^{-1} ({\bf q}, \omega)$,
for the comparison of the calculated loss-function spectra with experiment we refer to Ref. \onlinecite{Fugallo2015}. 
In the construction of the BSE hamiltonian, we expand the single-particle states and static dielectric function with plane-wave cutoffs up to 387 and 133 eV, 
respectively.  
We perform the KS and static screening calculations using ABINIT \cite{Abinit}, and BSE calculations with EXC \cite{EXC}. 
All the spectra presented in the following sections are calculated for in-plane momentum transfer $\bfq$ along the $\Gamma M$ direction. 

\section{Results}
\label{sec:results}

\begin{figure}
\includegraphics[width=\columnwidth, angle=0]{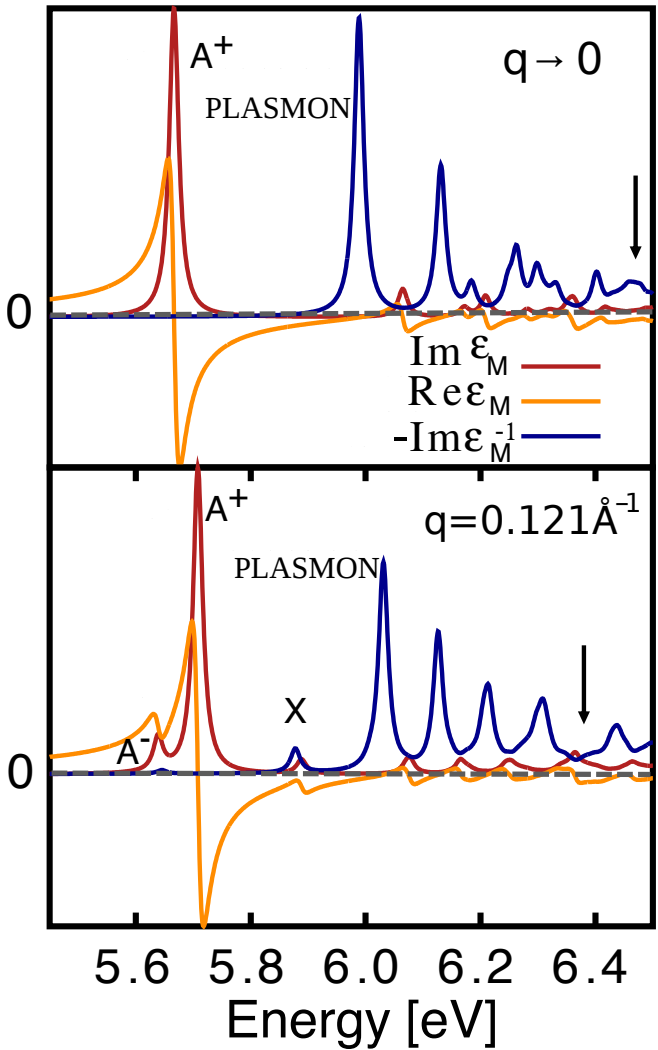}
\begin{center}
\caption{The real and imaginary parts of the macroscopic dielectric function $\textrm{Re} \epsilon_M$ and $\textrm{Im} \epsilon_M$
and the loss function $-\textrm{Im} \epsilon_M^{-1}$ 
calculated at two different wave vectors $\bfq$ along the in-plane $\Gamma M$ direction. 
For improved visibility the loss functions have been rescaled. The vertical arrows mark the smallest independent-particle GW transition energy
(which for $\bfq=0$ corresponds to the direct band gap).}
\label{fig1}
\end{center}
\end{figure}

The two panels of Fig. \ref{fig1} display the real and imaginary parts of macroscopic dielectric function $\epsilon_M$ 
and the loss function $-\rm{Im} \epsilon_M^{-1}$ of the bulk crystal of hBN calculated by solving the BSE for two different in-plane momenta $\bfq$. 
At vanishing $\bfq$ (top panel of Fig. \ref{fig1}) the prominent peak at 5.67 eV in the absorption spectrum $\rm{Im} \epsilon_M$
is a tightly bound exciton, located well within the direct band gap\footnote{In hBN the fundamental band gap is indirect \protect\cite{Arnaud2006} and in GW it is 5.78 eV.} (which in GW amounts to 6.47 eV and is marked by the vertical arrow in the top panel of Fig. \ref{fig1}).
In the plot we have labeled the main peak as ``A$^+$'' (the explanation of the identification of the various excitations will be the subject of the detailed analysis in Sec. \ref{sec:discussion}).
Other structures are visible in the spectrum at higher energies, but for simplicity here and in the following  we will focus on the lowest-energy excitations.
As explained in previous works \cite{Arnaud2006,Wirtz2006,Arnaud2008,Wirtz2008},
the main absorption peak derives from $\pi-\pi^*$ transitions between top-valence and bottom-conduction bands that are visible for in-plane light 
polarization \cite{Doni1969}.
Through the Kramers-Kronig relation, this  ``A$^+$'' peak of $\rm{Im} \epsilon_M$ induces a strong oscillation in  $\rm{Re} \epsilon_M$, which crosses the zero axis with a positive slope at 5.99 eV.
$\rm{Im} \epsilon_M$ being small at this energy, this zero of $\rm{Re} \epsilon_M$ gives rise to a plasmon resonance in the loss function $-\rm{Im} \epsilon_M^{-1}$, 
which shows a peak at the same energy [see Eq. \eqref{lossf}].
It is here worth noticing that in hBN also this plasmon excitation lies within the direct gap, 
since the collective charge excitation of the $\pi$ electrons is strongly affected by the e-h attraction\cite{Fugallo2015}.
As discussed in details in Refs. \onlinecite{Galambosi2011,Fugallo2015},
for increasing $\bfq$ this $\pi$ plasmon
disperses to higher energies and at larger $\bfq$ it enters the continuum of particle-hole excitations.

As a matter of example, the bottom panel of Fig. \ref{fig1} shows the  spectra obtained for the second smallest finite $\bfq$ that we have considered in our calculations 
(for the other momentum transfers, not shown here, similar considerations can be made). 
Globally the spectra at finite $\bfq$ remain qualitatively similar to the $\bfq=0$ case shown in the top panel of Fig. \ref{fig1}.
Still we can recognize that in $\rm{Im} \epsilon_M$ a new small structure ``A$^-$'' appears on  the low-energy side of the main ``A$^+$'' peak.
The ``A$^-$'' feature also induces a new shoulder in $\rm{Re} \epsilon_M$ in the same energy range.
Finally, a new small peak ``X'' is visible in the loss function $-\rm{Im} \epsilon_M^{-1}$ at 5.88 eV, i.e. before the $\pi$ plasmon. 
This new peak (which does not take place in correspondence with a zero of $\rm{Re} \epsilon_M$, hence it is not a plasmon) 
matches a new very weak peak in $\rm{Im} \epsilon_M$, so it has to be ascribed to a new many-body electron-hole excitation that becomes active at $\bfq\neq0$.

\begin{figure*}[ht]
\includegraphics[width= \textwidth, angle=0]{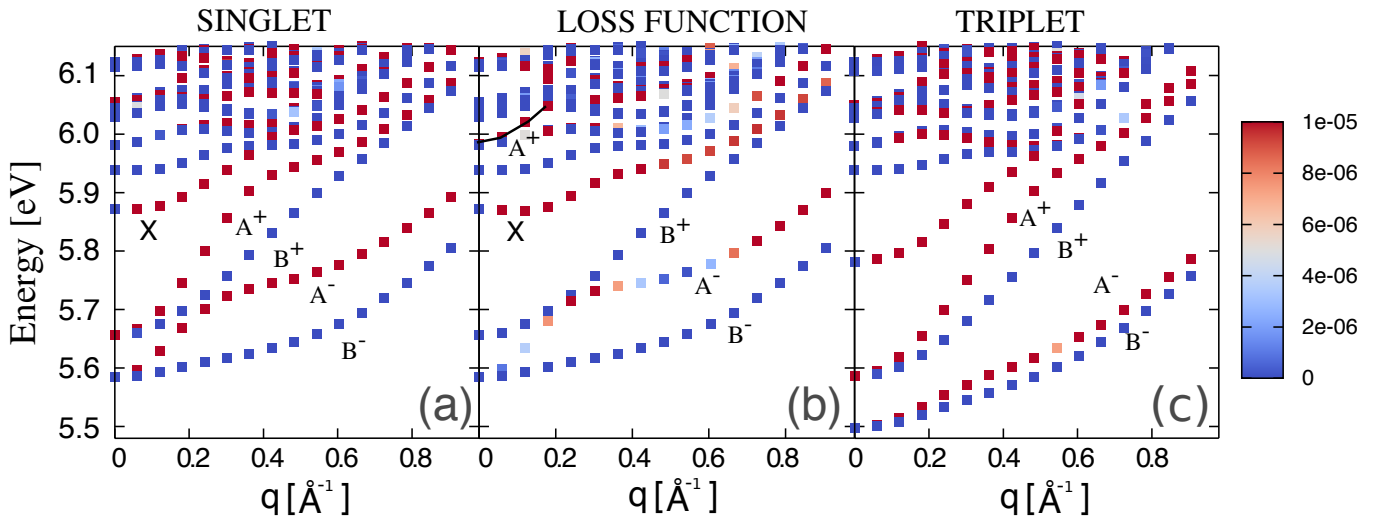}
\begin{center}
\caption{Exciton eigenvalue spectrum $E^\lambda(\bfq$) in bulk hBN for in-plane $\bfq$ along $\Gamma$M for (a) the electron-hole correlation function $\bar{L}$ corresponding to $\textrm{Im} \epsilon_M(\bfq,\w)$ 
featuring singlet excitons, 
(b)  the electron-hole correlation function $L$ corresponding to the loss function $-\textrm{Im} \epsilon_M^{-1}(\bfq,\w)$, displaying plasmons and e-h excitations, and (c) for triplet excitons. The $\Gamma$M length is 1.45 \AA$^{-1}$.
For the explanation of the peak labels see the main text.  In the loss-function plot, panel (b), the solid black line is a guide for the eye in order to better track the plasmon dispersion (corresponding to the A$^+$ feature).}
\label{fig2}
\end{center}
\end{figure*}

\begin{figure*}[ht]
\includegraphics[width= \textwidth, angle=0]{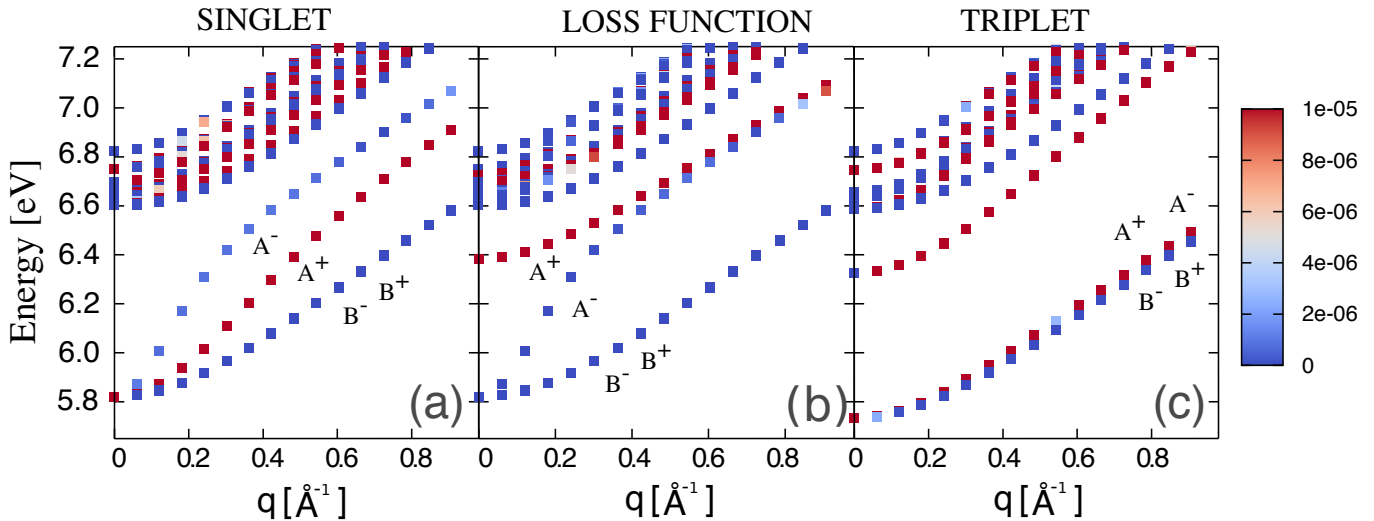}
\begin{center}
\caption{Same as Fig. \protect\ref{fig2} for increased interlayer distance $d=1.5d_0$ (where $d_0$ is the experimental interlayer separation of hBN).}
\label{fig4}
\end{center}
\end{figure*}

As discussed in Sec. \ref{sec:theory}, 
the spectra for $\rm{Im} \epsilon_M$ and $-\rm{Im} \epsilon_M^{-1}$ can be also 
analysed in detail  by making use of the eigenvalues and eigenvectors of the excitonic hamiltonian that enter Eq. \eqref{epm} and Eq. \eqref{epm-1}, respectively.
Fig. \ref{fig2}(a) shows the 18 lowest energies $E^{\lambda}$ as a function of $\bfq$ for the singlet excitons that are
obtained from the diagonalization of excitonic hamiltonian $H_{ex}$ \eqref{eqhex}. 
They are  hence the poles of $\epsilon_M$ \eqref{epm} and of the modified two-particle correlation function $\bar L$.
The color scale represents their intensity $|A^\lambda \cdot \tilde \rho|$  at the numerator of Eq. \eqref{epm}.
Red squares are for states that have a visible peak in $\rm{Im} \epsilon_M$, while blue squares are dark exciton states with no intensity in the spectrum.
The other two panels of Fig. \ref{fig2} use the same representation. 
Fig. \ref{fig2}(b) displays the exciton eigenvalues $E'_\lambda(\bfq)$ obtained from the diagonalization of $H_{ex}'$ 
that includes the long-range Coulomb interaction: they enter the loss function spectra  $-\rm{Im} \epsilon_M^{-1}(\bfq,\w)$ \eqref{epm-1}.
Finally, in Fig. \ref{fig2}(c) the triplet exciton energies are also reported for comparison (they cannot be directly   measured by loss or absorption spectroscopies). 
They are calculated from the excitonic hamiltonian $H_{ex}$ \eqref{eqhex} where the e-h exchange interaction $\bar v$ is absent.
With respect to the singlet excitons the triplet energies are globally lower [compare Fig. \ref{fig2}(a) and \ref{fig2}(c)], 
as the e-h exchange interaction is repulsive and hence yields singlet states that have higher energies than the corresponding triplets.

The first and third $\bfq$ points  in Figs. \ref{fig2}(a)-\ref{fig2}(b) allow us to understand better the absorption and loss spectra plotted 
in the two panels of Fig. \ref{fig1}.
For example, in Fig. \ref{fig2}(a) 
we discover that in the optical limit $\bfq\rightarrow0$ the first visible exciton ``A$^+$'' is degenerate with a dark state \footnote{Note that for degenerate eigenstates, 
the numerical diagonalisation of the excitonic hamiltonian  in principle can give as a result any linear combination of them.} (labelled ``B$^+$'' here)
and that below them there are other two degenerate dark excitons ``A$^-$'' and ``B$^-$''  that  do not contribute to the $\bfq\rightarrow0$ 
absorption spectrum in the top panel of Fig. \ref{fig1} (this point was already subject of discussion in Refs. \onlinecite{Arnaud2006,Wirtz2008,Arnaud2008}).
We can also see that at finite $\bfq$ one of the two lowest dark excitons becomes visible, giving rise to the low-energy peak ``A$^-$'' in the absorption 
spectrum of Fig. \ref{fig1}, bottom panel. Finally, the weak peak ``X'' at 5.88 eV is due to another exciton state that is dark at $\bfq=0$ and switches on at $\bfq\neq0$.
For all wavevectors $\bfq$, at higher energies the exciton states become very dense, forming a continuum of excitations. 
In Sec. \ref{sec:discussion} we will focus on the 4 lowest-energy discrete states that are well within the fundamental gap.

We can now repeat the same analysis for the loss functions $-\rm{Im} \epsilon_M^{-1}(\bfq,\w)$ in Fig. \ref{fig1} using the poles of $L$ represented in Fig. \ref{fig2}(b).
We thus discover that at  $\bfq = 0$ the plasmon excitation at 5.99 eV is not the lowest-energy eigenvalue. 
It  is actually located already in the energy region where e-h excitations are rather dense. So it is not easy to track its dispersion after the first few $\bfq$ points.
At the bottom of the eigenvalue spectrum  there are instead 3 dark states (2 of them are degenerate at $\bfq=0$) that are well separated from the other excitations.
They have a  dispersion as a function of $\bfq$  that is similar to that of the lowest poles of $\bar L$ in Fig. \ref{fig2}(a).
It is hence tempting to make a connection between them. In Sec. \ref{sec:discussion} we will explain rigorously why this is indeed the case 
(so they are labeled  ``A$^-$'' and ``B$^\pm$'' here)  and why the plasmon excitation instead has a ``A$^+$'' character.
Finally, at $\bfq=0$ at 5.88 eV  we recognize the same ``X'' excitation that is present also in the spectrum of $\bar L$ in Fig. \ref{fig2}(a) and is responsible for the weak structures in the absorption and loss
spectra in the bottom panel of Fig. \ref{fig1}.

\section{Discussion}
\label{sec:discussion}

\subsection{The exciton hamiltonian in layered crystals}

In order to  interpret  the numerical results of the previous section, 
here we generalize the approach that some of us introduced in Ref. \onlinecite{Cudazzo2012} to explain 
the excitonic properties of molecular crystals.
We thus rewrite the excitonic hamiltonian $\hat{H}_{ex}$ 
[which in Eq. \eqref{eqhex} is expressed in terms of Bloch wave functions delocalised all over the crystal]
in the basis of wave functions localized on the elementary units of the system. 
While in molecular crystals the elementary units are the single molecules, 
in the present case they are the single layers of BN (stacked along the $z$ axis).
We assume that the one-particle wave functions $\psi(\mathbf{r})$ localized on different layers 
do not overlap and can be factorized 
in an in-plane $\phi(\boldsymbol{\rho})$ and out-of-plane $\chi(z)$ components, with  $\bfr=(\boldsymbol{\rho},z)$.
Specifically, for given in-plane wave vector $\mathbf{k}$  and out-of-plane $k_z$, 
the single-particle wave function $\psi_{\mathbf{k},k_z}(\mathbf{r})$ is expanded in the basis of single-layer wave functions as: 
$\sum_{\mathbf{R}i} c_{\mathbf{k}}^i e^{ik_z\mathbf{R}} \phi^i_{\mathbf{k}^{(i)}}(\boldsymbol{\rho})\chi^i(z-\mathbf{R})$. 
Here $\mathbf{R}$ is the lattice vector along $z$ and the index $i$ denotes the layers inside the unit cell. 
We also consider the possibility that the various layers stacked along $z$ are rotated one with respect to another by an angle $\beta$ (in hBN $\beta=60^{\circ}$), 
and therefore also the 2D first Brillouin zones are rotated by an angle $\beta$\cite{*[{The effect of stacking order in hBN has been investigated within the BSE in: }][{}] Bourrellier2014}.
Hence, choosing a reference layer $i=1$, we define $\mathbf{k}^{(i)}=\mathbf{k}$ for $i=1$ and $\mathbf{k}^{(i)}=\boldsymbol{\beta}^{-1}\mathbf{k}$ for $i\neq1$,
$\boldsymbol{\beta}\mathbf{k}$ being the wave vector obtained rotating $\mathbf{k}$ by an angle $\beta$ (see App. \ref{app:a} for more details).
For simplicity we further consider for each layer a two-bands system, with only one valence $v$ and one conduction $c$ bands.
Under these assumptions, 
the whole excitonic hamiltonian $\hat{H}_{ex}$ of the crystal Eq. \eqref{eqhex}
takes the simple form of the sum of three terms 
$\hat{H}_{ex}= \hat H_{ip} + \hat K_{CT}+ \hat K_{FR}$:
\begin{widetext}
\begin{align}\label{eq1}
\hat{H}_{ip} =& \sum_{\mathbf{k}_1\mathbf{k}_2}\sum_{\mathbf{R}\mathbf{S}}\sum_{ij}E^{\mathbf{R}i\mathbf{S}j}_c(\mathbf{k}_1,\mathbf{k}_2)a^{\dag}_{c\mathbf{k}_1\mathbf{R}i}a_{c\mathbf{k}_2\mathbf{S}j}-\sum_{\mathbf{k}_1\mathbf{k}_2}\sum_{\mathbf{R}\mathbf{S}}\sum_{ij}E^{\mathbf{R}i\mathbf{S}j}_v(\mathbf{k}_1,\mathbf{k}_2)b^{\dag}_{v\mathbf{k}_1\mathbf{R}i}b_{v\mathbf{k}_2\mathbf{S}j} \nonumber \\ 
\hat{K}_{FR}=& \sum_{\mathbf{k}_1\mathbf{k}_2\mathbf{k}_3\mathbf{k}_4}\sum_{\mathbf{R}i,\mathbf{S}j}
\left[ \bar v^{\mathbf{S}j,\mathbf{S}j}_{\mathbf{R}i,\mathbf{R}i}(v\mathbf{k}_1c\mathbf{k}_2v\mathbf{k}_3c\mathbf{k}_4)-\delta_{\mathbf{R}i,\mathbf{S}j}W^{\mathbf{R}i,\mathbf{R}i}_{\mathbf{R}i,\mathbf{R}i}(v\mathbf{k}_1c\mathbf{k}_2v\mathbf{k}_3c\mathbf{k}_4) \right]
a^{\dag}_{c\mathbf{k}_2\mathbf{R}i}b^{\dag}_{v\mathbf{k}_1\mathbf{R}i}b_{v\mathbf{k}_3\mathbf{S}j}a_{c\mathbf{k}_4\mathbf{S}j} \nonumber \\
\hat{K}_{CT}=&
-\sum_{\mathbf{k}_1\mathbf{k}_2\mathbf{k}_3\mathbf{k}_4}\sum_{\mathbf{R}i,\mathbf{S}j}(1-\delta_{\mathbf{R}i,\mathbf{S}j})W^{\mathbf{S}j,\mathbf{R}i}_{\mathbf{S}j,\mathbf{R}i}(v\mathbf{k}_1c\mathbf{k}_2v\mathbf{k}_3c\mathbf{k}_4)
a^{\dag}_{c\mathbf{k}_2\mathbf{R}i}b^{\dag}_{v\mathbf{k}_1\mathbf{S}j}b_{v\mathbf{k}_3\mathbf{S}j}a_{c\mathbf{k}_4\mathbf{R}i},
\end{align}
with
\begin{align}
W^{\mathbf{S}j,\mathbf{R}i}_{\mathbf{S}j,\mathbf{R}i}(v\mathbf{k}_1c\mathbf{k}_2v\mathbf{k}_3c\mathbf{k}_4)  = & \int d\mathbf{r}d\mathbf{r}'\phi^{i*}_{c\mathbf{k}_2}(\boldsymbol{\rho})\chi^{i*}_c(z-\mathbf{R})\chi^{j*}_{v}(z'-\mathbf{S})\phi^{j*}_{v\mathbf{k}_3}(\boldsymbol{\rho}')W(\mathbf{r},\mathbf{r}') \phi^{i}_{c\mathbf{k}_4}(\boldsymbol{\rho})\chi^{i}_{c}(z-\mathbf{R})\phi^{j}_{v\mathbf{k}_1}(\boldsymbol{\rho}')\chi^{j}_v(z'-\mathbf{S}) \\
\bar v^{\mathbf{S}j,\mathbf{S}j}_{\mathbf{R}i,\mathbf{R}i}(v\mathbf{k}_1c\mathbf{k}_2v\mathbf{k}_3c\mathbf{k}_4)  = & \int d\mathbf{r}d\mathbf{r}'\phi^{i*}_{c\mathbf{k}_2}(\boldsymbol{\rho})\chi^{i*}_c(z-\mathbf{R})\phi^{j*}_{v\mathbf{k}_3}(\boldsymbol{\rho}')\chi^{j*}_{v}(z'-\mathbf{S})\bar{v}(\mathbf{r},\mathbf{r}') \phi^{j}_{c\mathbf{k}_4}(\boldsymbol{\rho}')\chi^{j}_{c}(z'-\mathbf{S})\phi^{i}_{v\mathbf{k}_1}(\boldsymbol{\rho})\chi^{i}_v(z-\mathbf{R}).
\end{align}
\end{widetext}

In the Bloch picture $\hat H_{ip}$ contains independent e-h transitions between  single-particle bands.
Equivalently, here $\hat H_{ip}$  describes scattering processes  from layer to layer, independently for electrons and holes, being
\beq
E^{\mathbf{R}i\mathbf{S}j}_{v(c)}(\mathbf{k}_1,\mathbf{k}_2)=E^i_{v(c)}(\mathbf{k}_1)\delta_{\mathbf{R}i,\mathbf{S}j}\delta_{\mathbf{k}_1,\mathbf{k}_2}+t^{v(c)\mathbf{k}_1,\mathbf{k}_2}_{\mathbf{R}i,\mathbf{S}j},
\label{eq:hop}
\eeq
 where $E^i_{v(c)}(\mathbf{k}_1)$ is the single-layer band dispersion 
 and $t^{v(c)\mathbf{k}_1,\mathbf{k}_2}_{\mathbf{R}i,\mathbf{S}j}$ 
 are interlayer hopping matrix elements (see App. \ref{app:a}) 
 that give rise to the finite $k_z$  dispersion of the bands in the crystal (see Fig. \ref{fig8} in App. \ref{app:c}).
 
 In Eq. \eqref{eq1} the second and third terms $\hat{K}_{FR}$ and $\hat{K}_{CT}$  describe the interaction between an electron and a hole that are localized on the same layer or on different layers, respectively.
 In order to keep a closer contact with the exciton physics of molecular crystals, 
 here we name the intralayer configuration as a ``Frenkel'' (FR) exciton and the interlayer configuration as a ``charge-transfer'' (CT) exciton. In other words, in the present context we call FR an exciton that is fully localised on a single layer, 
 independently of being localised or not within the layer. 
 Therefore this definition applies equivalently for excitons with different in-plane localisation characters, as for example in hBN (where the exciton is tightly bound also within the single layer \cite{Arnaud2006}) or in MoS$_2$ (where it is weakly bound\cite{Molina2013}).
 We note that 
 the e-h exchange interaction $\bar v$ is different from zero only for e-h pairs localized on the same layer, therefore it is absent for CT excitons in Eq. \eqref{eq1}.
 
 The FR and CT interaction terms in Eq. \eqref{eq1} are coupled by the interlayer hopping terms in $\hat{H}_{ip}$. 
 Without  the interlayer hopping $t^{v(c)\mathbf{k}_1,\mathbf{k}_2}_{\mathbf{R}i,\mathbf{S}j}$
 the excitonic hamiltonian \eqref{eq1} factorizes into two independent blocks: 
 a CT hamiltonian $\hat H_{CT}= \hat H_{ip}^\prime+ \hat K_{CT}$  
 describing an interacting e-h pair localised on different layers and 
 a FR hamiltonian $\hat H_{FR}= \hat H_{ip}^\prime+ \hat K_{FR}$ describing an interacting e-h pair on the same layer (in both cases we set $t^{v(c)\mathbf{k}_1,\mathbf{k}_2}_{\mathbf{R}i,\mathbf{S}j}=0$ in $H_{ip}^\prime$).
 
The CT exciton wave functions\footnote{Charge-transfer excitons in transition-metal dichalcogenide heterobilayer have been recently investigated using a Wannier model e.g. in Refs. \protect\onlinecite{Yu2015,Rivera2016}.}: 
\beq\label{eq:CT}
|\Psi^{CT}_{ex}(\mathbf{q})\rangle = \sum_{\lambda} \sum_{ij\tau} c_{ij}^{\tau}|\Psi^{\lambda}_{ij,\boldsymbol{\tau}}(\mathbf{q})\rangle
\eeq
with
 \beq 
 |\Psi^{\lambda}_{ij,\boldsymbol{\tau}}(\mathbf{q})\rangle=\frac{1}{\sqrt{N}}\sum_{\mathbf{R}} \sum_{\mathbf{k}}A^{\lambda,ij,\boldsymbol{\tau}}_{vc,\mathbf{k}}(\mathbf{q})a^{\dag}_{c\mathbf{k}^{(i)}\mathbf{R}i}b^{\dag}_{v\mathbf{k}^{(j)}+\mathbf{q}^{(j)}\mathbf{R}+\boldsymbol{\tau}j}|0\rangle
 \eeq
are already the eigenfunctions of $\hat{H}_{CT}$ that can be directy built from the excitations of the single layers. 
The Frenkel hamiltonian $\hat{H}_{FR}$ instead contains also an interlayer coupling that needs additional consideration. 

By further splitting the e-h exchange interaction $\bar v$ into a long-range contribution  $\bar{v}_0$
(corresponding to the $\mathbf{G}_{||}=0$ component in reciprocal space) and a 
short-range contribution $\bar{\bar{v}}$ such that $\bar v = \bar{v}_0 + \bar{\bar{v}}$, 
the FR hamiltonian $\hat H_{FR}$ can be separated into an intralayer term $\hat{H}_L$ and an interlayer coupling $\hat{\bar V}$, 
$\hat H_{FR}=\hat{H}_L+ \hat{\bar V}$, with:
\begin{widetext}
\begin{align}
\hat{H}_L =&  
H_{ip}^\prime 
+ \sum_{\mathbf{k}_1\mathbf{k}_2\mathbf{k}_3\mathbf{k}_4}\sum_{\mathbf{R}i} \left[\bar{\bar{v}}^{\mathbf{R}i,\mathbf{R}i}_{\mathbf{R}i,\mathbf{R}i}
(v\mathbf{k}_1c\mathbf{k}_2v\mathbf{k}_3c\mathbf{k}_4)-W^{\mathbf{R}i,\mathbf{R}i}_{\mathbf{R}i,\mathbf{R}i}(v\mathbf{k}_1c\mathbf{k}_2v\mathbf{k}_3c\mathbf{k}_4)\right]a^{\dag}_{c\mathbf{k}_2\mathbf{R}i}b^{\dag}_{v\mathbf{k}_1\mathbf{R}i}b_{v\mathbf{k}_3\mathbf{R}i}a_{c\mathbf{k}_4\mathbf{R}i} \\
\hat{\bar V} = &\sum_{\mathbf{k}_1\mathbf{k}_2\mathbf{k}_3\mathbf{k}_4}\sum_{\mathbf{R}i,\mathbf{S}j}\bar v^{\mathbf{S}j,\mathbf{S}j}_{0\mathbf{R}i,\mathbf{R}i}(v\mathbf{k}_1c\mathbf{k}_2v\mathbf{k}_3c\mathbf{k}_4)a^{\dag}_{c\mathbf{k}_2\mathbf{R}i}b^{\dag}_{v\mathbf{k}_1\mathbf{R}i}b_{v\mathbf{k}_3\mathbf{S}j}a_{c\mathbf{k}_4\mathbf{S}j}
\end{align}
where
\begin{align}
\bar{\bar{v}}^{\mathbf{S}j\mathbf{S}j}_{\mathbf{R}i\mathbf{R}i}(v\mathbf{k}_1c\mathbf{k}_2v\mathbf{k}_3c\mathbf{k}_4) =& \sum_{\mathbf{q}_{||}}\delta_{\mathbf{k}_1,\mathbf{k}_2+\mathbf{q}^{(i)}_{||}}\delta_{\mathbf{k}_3,\mathbf{k}_4+\mathbf{q}^{(j)}_{||}}
\sum_{q_zG_z}\sum_{\mathbf{G}_{||}\neq 0} \frac{4\pi}{|\mathbf{q}+\mathbf{G}|^2}\tilde{\rho}^i_{c\mathbf{k}_2v\mathbf{k}_1}(\mathbf{q}^{(i)}_{||}+\mathbf{G}^{(i)}_{||})\tilde{\rho}^{j*}_{c\mathbf{k}_4v\mathbf{k}_1}(\mathbf{q}^{(j)}_{||}+\mathbf{G}^{(j)}_{||}) \nonumber \\
& \times
\left|\int dz\chi^*_c(z)e^{i(q_z+G_z)z}\chi_v(z)\right|^2e^{-iq_z\cdot(\mathbf{S}-\mathbf{R})}e^{-i(q_z+G_z)d_{ij}}  \label{eqbarv} \\
\bar v^{\mathbf{S}j\mathbf{S}j}_{0\mathbf{R}i\mathbf{R}i}(v\mathbf{k}_1c\mathbf{k}_2v\mathbf{k}_3c\mathbf{k}_4) =& \sum_{\mathbf{q}_{||}}\delta_{\mathbf{k}_1,\mathbf{k}_2+\mathbf{q}^{(i)}_{||}}\delta_{\mathbf{k}_3,\mathbf{k}_4+\mathbf{q}^{(j)}_{||}}\sum_{\substack 
{q_z,  G_z\neq0, \\ \mathbf{G}_{||}=0}}\frac{4\pi}{|\mathbf{q}+\mathbf{G}|^2}\tilde{\rho}^i_{c\mathbf{k}_2v\mathbf{k}_1}(\mathbf{q}^{(i)}_{||})\tilde{\rho}^{j*}_{c\mathbf{k}_4v\mathbf{k}_1}(\mathbf{q}^{(j)}_{||}) \nonumber \\
&\times
\left|\int dz\chi^*_c(z)e^{i(q_z+G_z)z}\chi_v(z)\right|^2e^{-iq_z\cdot(\mathbf{S}-\mathbf{R})}e^{-i(q_z+G_z)d_{ij}}  \label{eqv0} \\
\tilde{\rho}^i_{c\mathbf{k}v\mathbf{k}'}(\mathbf{q}^{(i)}_{||}+\mathbf{G}^{(i)}_{||})= &
\int d\boldsymbol{\rho}\phi^{i*}_{c\mathbf{k}}(\boldsymbol{\rho})e^{i(\mathbf{q}^{(i)}_{||}+\mathbf{G}^{(i)}_{||})\cdot\boldsymbol{\rho}}\phi^{i}_{v\mathbf{k}'}(\boldsymbol{\rho}).
\end{align}
Equivalenty, Eq. \eqref{eqbarv} 
can be written in terms of the partial Fourier transform of the Coulomb potential\cite{Hambachphd}:
\beq
v(\mathbf{q}_{||}+\mathbf{G}_{||},z,z') = \frac{2\pi}{|\mathbf{q}_{||}+\mathbf{G}_{||}|}e^{-|\mathbf{q}_{||}+\mathbf{G}_{||}||z-z'|}
\eeq
as:
\begin{align}
\bar{\bar{v}}^{\mathbf{S}j\mathbf{S}j}_{\mathbf{R}i\mathbf{R}i}(v\mathbf{k}_1c\mathbf{k}_2v\mathbf{k}_3c\mathbf{k}_4) =& \sum_{\mathbf{q}_{||}}\delta_{\mathbf{k}_1,\mathbf{k}_2+\mathbf{q}^{(i)}_{||}}\delta_{\mathbf{k}_3,\mathbf{k}_4+\mathbf{q}^{(j)}_{||}}
\sum_{\mathbf{G}_{||}\neq 0} \frac{2\pi}{|\mathbf{q}_{||}+\mathbf{G}_{||}|}\tilde{\rho}^i_{c\mathbf{k}_2v\mathbf{k}_1}(\mathbf{q}^{(i)}_{||}+\mathbf{G}^{(i)}_{||})\tilde{\rho}^{j*}_{c\mathbf{k}_4v\mathbf{k}_1}(\mathbf{q}^{(j)}_{||}+\mathbf{G}^{(j)}_{||}) \nonumber \\
& \times
\int dz\int dz'\chi^*_c(z)\chi_v(z)e^{-|\mathbf{q}_{||}+\mathbf{G}_{||}||z-z'|}\chi^*_v(z')\chi_c(z')e^{-|\mathbf{q}_{||}+\mathbf{G}_{||}||\mathbf{S}-\mathbf{R}|}e^{-|\mathbf{q}_{||}+\mathbf{G}_{||}|d_{ij}}.  \label{eqbarv2} 
\end{align}
\end{widetext}
From  Eq. \eqref{eqbarv2} we can conclude that the
off-diagonal elements  $\mathbf{S} \neq \mathbf{R}$ and $i \neq j$  of $\bar{\bar{v}}^{\mathbf{S}j\mathbf{S}j}_{\mathbf{R}i\mathbf{R}i}$ are actually zero,
for the presence of the exponential terms $e^{-|\mathbf{q}_{||}+\mathbf{G}_{||}||\mathbf{S}-\mathbf{R}|}$ with $\mathbf{G}_{||}\neq 0$.
For its short-range nature, the $\bar{\bar{v}}$ interaction therefore does not couple different layers. 

With respect to the Bloch picture, such a transformation and decomposition of the excitonic hamiltonian \eqref{eqhex}
illustrates much more clearly the physics of excitons in layered materials that we want to uncover.
Here the eigenstates of $\hat{H}_L$ represent the excitations of an elementary unit of our van der Waals material, 
namely a single BN layer embedded in the bulk crystal.
They are formally analogous to the excitations of a single molecule in a molecular solid. 
Thus, by analogy with molecular crystals, a FR exciton in the present case can be seen as an elementary excitation of a single layer, 
which can scatter from one layer to another due to the interlayer coupling $\hat{\bar V}$.
From a mathematical point of view, this means that
we expand the FR exciton wave functions 
(which are the eigenfunctions of  $\hat{H}_{FR}$) on the basis of the eigenstates of $\hat{H}_L$: 
\beq
|\Psi^{FR}_{ex}(\mathbf{q})\rangle=\frac{1}{\sqrt{N}}\sum_{\lambda,\mathbf{R}i}c^{\lambda}_i(\mathbf{q})|\Psi^{\lambda}_{\mathbf{R},ii}(\mathbf{q})\rangle
\eeq
where
\beq
|\Psi^{\lambda}_{\mathbf{R},ii}(\mathbf{q})\rangle=\sum_{\mathbf{k}}A^{\lambda,i}_{vc,\mathbf{k}}(\mathbf{q})a^{\dag}_{c\mathbf{k}^{(i)}\mathbf{R}i}b^{\dag}_{v\mathbf{k}^{(i)}+\mathbf{q}^{(i)}\mathbf{R}i}|0\rangle.
\eeq
and where we have used the fact that for in-plane $\bfq$ $e^{i\mathbf{q} \cdot \mathbf{R} } = 1$.
The matrix elements of $\hat{\bar V}$  are: 
 \begin{multline} \label{Vmat}
\langle\Psi^{\lambda}_{\mathbf{R},ii}(\mathbf{q})|\hat{\bar V}|\Psi^{\lambda'}_{\mathbf{R}',jj}(\mathbf{q})\rangle=\sum_{G_z \neq 0,\mathbf{G}_{||}=0}\frac{4\pi}{|\mathbf{q}+\mathbf{G}|^2}  \\ 
\times [S_i^\lambda(\bfq)]^* [S_j^{\lambda'}(\bfq)]
\left|\int dz\chi^*_c(z)e^{iG_zz}\chi_v(z)\right|^2e^{-iG_zd_{ij}},
\end{multline}
where $S_i^\lambda(\bfq)$ is the oscillator strength of the 
 exciton $\lambda$ of the layer $i$:
\beq
S_i^\lambda(\bfq)=\sum_{\bfk} A^{\lambda,i}_{vc,\mathbf{k}}(\mathbf{q}^{(i)}) \tilde{\rho}^i_{vc\mathbf{k}}(\mathbf{q}^{(i)}),
\eeq
and
where we have used the fact that $\chi_{n}^i(z)=\chi_{n}^j(z-d_{ij})\equiv\chi_{n}(z-d_{ij})$ for both $n=v,c$, 
with $d_{ij}$ the distance between the layers $i$ and $j$. 
From Eq. \eqref{Vmat} we realise that $\hat{\bar V}$ operates only on visible excitons and cannot couple visible and dark excitons for which $S_i^\lambda(\bfq)=0$.

\subsection{The exciton hamiltonian in hBN}

If we consider a crystal with two inequivalent layers per unit cell, as it is the case for hBN,
for each quantum number $\lambda$ that defines an excitation of the single layer one has four excitons in the bulk [we take into account only first nearest-neighbor CT excitons and assume $\boldsymbol{\tau}=0$ in Eq. \eqref{eq:CT}].
The FR and CT excitons that  diagonalize the excitonic hamiltonian $\hat{H}_{ex}$ \eqref{eq1}  in absence of interlayer hopping
are then the symmetric and antisymmetric combinations with respect to the exchange of the e-h pair between two inequivalent layers:
\begin{align}
|CT^{\lambda}_{\pm}\rangle =& \frac{1}{\sqrt{N}}\sum_{\mathbf{R}}\frac{1}{\sqrt{2}} \left[|\Psi^{\lambda}_{\mathbf{R},12}\rangle\pm|\Psi^{\lambda}_{\mathbf{R},21}\rangle \right] \\
|FR^{\lambda}_{\pm}\rangle =& \frac{1}{\sqrt{N}}\sum_{\mathbf{R}}\frac{1}{\sqrt{2}}
\left[|\Psi^{\lambda}_{\mathbf{R},11}\rangle\pm|\Psi^{\lambda}_{\mathbf{R},22}\rangle \right].
\end{align}
The $|CT^{\lambda}_{\pm}\rangle$ states are degenerate, while the energy separation between the $|FR^{\lambda}_{\pm}\rangle$ states in the context of molecular crystals is usually called Davydov splitting \cite{Davydov1971}.

\begin{figure}
\includegraphics[width= 0.95\columnwidth, angle=0]{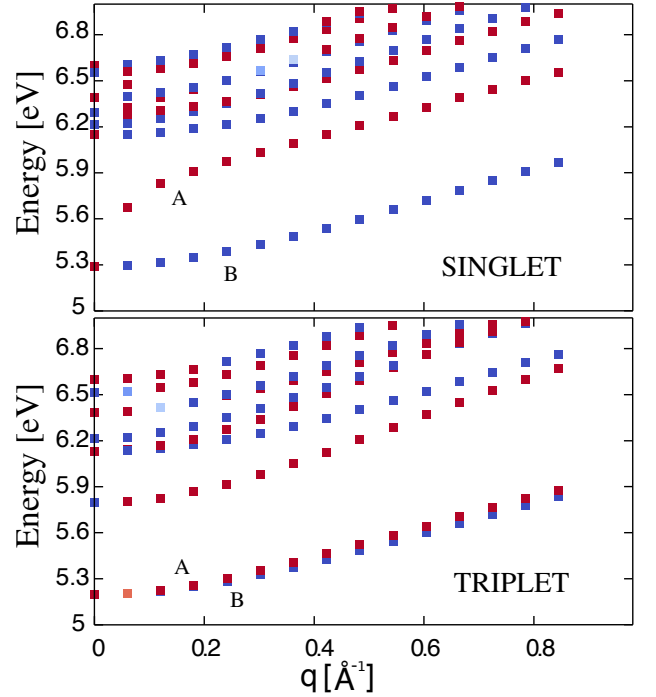}
\begin{center}
\caption{Dispersion of (a) singlet and (b) triplet excitons in the BN single layer. The color key is the same as in Fig. \protect\ref{fig2}.}
\label{fig3}
\end{center}
\end{figure}

In the case of hBN, the two lowest excitons of the BN monolayer, which are degenerate at $\bfq=0$, are
a visible exciton $A$ and a dark exciton $B$ [see Fig. \ref{fig3}(a)].
They originate from transitions from  the top-valence to the bottom-conduction bands with $\bfk$ vectors located around the $K$ or $K'$ points of the Brillouin zone, respectively\cite{Cudazzo2016,Galvani2016}.
These two intralayer A and B excitons hence produce eight excitons (four FR and four CT excitons) in the bulk crystal. 
Since the B exciton is dark for all $\bfq$ along $\Gamma$M, in the bulk the  A and B excitons are not mixed by $\hat {\bar V}$ [see Eq. \eqref{Vmat}] and preserve their identity.
The four $|CT^{A,B}_{\pm}\rangle$ excitons are located at higher energies since they have smaller binding energies, 
as a result of the e-h attraction being smaller for interlayer e-h pairs 
than for intralayer e-h pairs. 
In the following we focus on the four  $|FR^{A,B}_{\pm}\rangle$  excitons that are the lowest-energy excitations in the bulk.

The A exciton of the single layer of energy $E^A(\mathbf{q})=E_{11}^{A}(\bfq)=E_{22}^{A}(\bfq)$
gives rise to the two $|FR^{A}_{\pm}\rangle$  excitons:
\beq
E^A_{\pm}(\mathbf{q})=E^A(\mathbf{q})+\bar I^A(\mathbf{q})\pm \bar J^A(\mathbf{q})
\eeq
where  $\bar I^A(\mathbf{q})\pm \bar J^A(\mathbf{q})$ are the contribution to the exchange e-h interaction 
$\hat{\bar V}$ for the symmetric $(+)$ and antisymmetric $(-)$ states, respectively.
They are the excitation-transfer interactions that are responsible for the interlayer propagation of the FR exciton in the crystal\cite{Cudazzo2012}: 
$\bar J^A$ is related to the scattering process of an e-h pair
between two inequivalent layers and, analogously, $\bar I^A$ between equivalent layers in different unit cells.
Explicitly they read:
\begin{align}
\bar I^A(\mathbf{q}) =& \sum_{G_z\neq0, \mathbf{G}_{||}=0}\frac{4\pi}{|\mathbf{q}+\mathbf{G}|^2}  
|S^A(\bfq)|^2 \nonumber \\  
& \times \left|\int dz\chi^*_c(z)e^{iG_zz}\chi_v(z)\right|^2  \label{iqa} \\
\bar J^A(\mathbf{q}) =&\sum_{G_z\neq0,\mathbf{G}_{||}=0}\frac{4\pi}{|\mathbf{q}+\mathbf{G}|^2}   
|S^A(\bfq)|^2 \nonumber  \nonumber \\
& \times \left| \int dz\chi^*_c(z)e^{iG_zz}\chi_v(z) \right|^2e^{-iG_zd}. \label{jqa}
\end{align}
We note that  $\bar I^A(\mathbf{q})$ and $\bar J^A(\mathbf{q})$  
are both zero at $\bfq=0$, since  the oscillator strength  $S^A(\bfq)$ in the dipole limit $\bfq\rightarrow0$ is proportional to $\mathbf{q}\cdot\boldsymbol{\mu}^A$. 
Therefore $\bar I$ and $\bar J$ in layered systems do not yield any Davydov splitting between symmetric and antisymmetric excitons at $\bfq=0$, in contrast to the molecular crystal case\cite{Cudazzo2012}.
The matrix elements for $\mathbf{r}=(\mathbf{\boldsymbol{\rho}},0)$ are:
\begin{multline} 
\langle FR^{A}_{\pm}|e^{i \bfq \mathbf{r}}|0\rangle=  
|S^A(\bfq)|^2  \\ \label{eqmu}
\times \left[ \int dz\chi^*_c(z)\chi_v(z)\pm\int dz\chi^*_c(z)\chi_v(z) \right].
\end{multline}
The symmetric $|FR^{A}_{+}\rangle$ exciton is hence visible, while the antisymmetric $|FR^{A}_{-}\rangle$ exciton is dark, 
since the two integrals in Eq. \eqref{eqmu} exactly cancel in this case.
For the $B$ exciton the matrix element of $\hat{\bar V}$ is zero [since 
$S^B(\bfq)=0$ in Eq. \eqref{Vmat}]. 
Therefore the two $|FR^{B}_{\pm}\rangle$ excitons remain degenerate in the bulk:
\beq 
E^B_{\pm}(\mathbf{q})=E^B(\mathbf{q}).
\eeq
Moreover, as 
$S^B(\bfq)$ is zero, they are both dark [see Eq. \eqref{eqmu}].

In summary, by neglecting the interlayer hopping terms in the exciton hamiltonian \eqref{eq1}, we would expect that the two lowest A and B  excitons of the BN single layer
give rise to 3 FR dark excitons and 1 FR visible exciton in the bulk (together with CT excitons at high energies).

The effect of the hopping is, in general, to couple FR and CT excitons.
This coupling produces states with mixed character, FR+CT and CT+FR respectively, and modifies their energies. 
In hBN, as demonstrated in the App. \ref{app:a}, at $\mathbf{q}=0$
$\langle CT^{\lambda}_\pm|\hat{T}|FR^{\lambda}_\mp\rangle=0$:
excitons with different parities do not couple, giving rise to $|(FR+CT)^{\lambda}_\pm\rangle$ states with well defined parity.
Moreover, since  $\langle CT^{\lambda}_+|\hat{T}|FR^{\lambda}_+\rangle \neq \langle CT^{\lambda}_-|\hat{T}|FR^{\lambda}_-\rangle$ (see App. \ref{app:a}),
at $\mathbf{q}=0$  the hopping induces a finite Davydov splitting between symmetric and antisymmetric excitons.
Instead at finite $\mathbf{q}$ the various excitons formally lose their parity character as FR and CT states with different parities are generally allowed to mix together.

\subsection{Exciton dispersion: electron-hole exchange and interlayer hopping}

On the basis of the previous analysis, we can now examine in detail the properties of the four lowest-energy singlet excitons in hBN [see Fig. \ref{fig2}(a)].
In particular, we can understand the effect of e-h exchange by comparing singlet and triplet excitons [see Fig. \ref{fig2}(a) and (c)], because in the latter there is no e-h exchange.
Moreover, we can suppress also the interlayer hopping by artificially increasing the interlayer distance $d$. 
The singlet and triplet exciton band structures obtained with $d=1.5 d_0$, where $d_0$ is the experimental interlayer distance  of hBN, 
are displayed in Fig. \ref{fig4}(a) and (c).
With this increased separation between BN layers, the interlayer hopping is  reduced so much that the  $k_z$  dispersion of the top-valence and bottom-conduction 
single-particle bands becomes negligible (see App. \ref{app:c}).

At $\bfq=0$ the four lowest singlet excitons are grouped in two pairs [see Fig. \ref{fig2}(a)].
Since in the single layer the A and B excitons are degenerate at  $\bfq=0$ [see Fig. \ref{fig3}(a)] and the e-h exchange terms $\bar I(\mathbf{q=0})$ and $\bar J(\mathbf{q=0})$
are zero for all of them [see Eqs. \eqref{iqa}-\eqref{jqa}], the energy splitting between the two pairs must derive from the interlayer hopping 
(which we reasonably assume to be the same for A and B excitons).
At $\mathbf{q}=0$ the hopping conserves the parity character, removing the degeneracy between symmetric and antisymmetric states.
Indeed this energy splitting is present also for the triplet excitons [see Fig. \ref{fig2}(c)], whereas it becomes zero for an increased interlayer distance [see Fig. \ref{fig4}(a) and (c)].
Therefore we can conclude that at $\bfq=0$ the two excitons of the lowest pair, which are both dark, are the antisymmetric $|(FR+CT)^{A}_-\rangle$ and $|(FR+CT)^{B}_-\rangle$ states, while the two excitons of the other pair are the  symmetric $|(FR+CT)^{A}_+\rangle$  (which is visible) and $|(FR+CT)^{B}_+\rangle$ (which is dark).
For simplicity, in Figs. \ref{fig2}-\ref{fig4} we have labeled ``A$^\pm$'' and  ``B$^\pm$'' respectively the states $|(FR+CT)^{A}_\pm\rangle$ and $|(FR+CT)^{B}_\pm\rangle$.

Having established the character of the excitons at  $\bfq=0$, we can now track their dispersion as a function of $\bfq$.
The fact that one of the excitons of the lowest pair that is dark at $\bfq=0$ becomes visible at $\bfq \neq 0$ for both the singlet and triplet cases [see Fig. \ref{fig2}(a) and (c)]
is another effect of the interlayer hopping that at $\bfq\neq0$ mixes FR and CT states with different parities. 
This means that the parity is no more a good quantum number and the eigenstates of the excitonic hamiltonian are combinations of $|(FR+CT)_-\rangle$ and $|(FR+CT)_+\rangle$ states. In this way the dark exciton $|(FR+CT)^A_-\rangle$ is switched on by the effective coupling with the visible exciton $|(FR+CT)^A_+\rangle$.
Formally all the excitons lose their defined parity, 
but here for simplicity we still call them ``A$^\pm$'' (the two visible states) and ``B$^\pm$'' (the two dark states).

In order to infer the effect of the interlayer hopping on the exciton dispersion, 
we compare the behavior of the triplet excitons
in the bulk [see Fig. \ref{fig2}(c)], for the increased interlayer distance [see Fig. \ref{fig4}(c)]
and  in the monolayer [see Fig. \ref{fig3}(b)].
In the monolayer the A and B triplet excitons are almost degenerate: there is a tiny separation due to  the direct e-h attraction $W$\cite{Cudazzo2016}.
The same holds for $d=1.5 d_0$, where there is no effect of the interlayer hopping. In the bulk, instead, the hopping acts differently for the various excitons, 
giving rise to  a finite dispersion that removes the degeneracies. The energy level ordering remains the same for all $\bfq$. 
From the bottom to the top one has the following states: ``B$^-$'',``A$^-$'',``B$^+$'', and ``A$^+$''. 

The difference between the dispersions of the singlet and the triplet excitons in Fig. \ref{fig2}(a) and (c)
illustrates the role of the e-h exchange interaction in the bulk as a function of $\bfq$.
While the ``B$^\pm$'' excitons keep the same dispersion in the two channels  (as the e-h exchange $\bar I(\mathbf{q})=\bar J(\mathbf{q})=0$ for B excitons),
for the ``A$^\pm$'' excitons we observe that the effect of the e-h exchange is larger for small $\bfq$  
than for large $\bfq$, where the dispersion of singlet and triplet excitons tend to be the same,
being determined by the single-particle band dispersion only.

At increased interlayer distance $d=1.5d_0$, in contrast to the bulk,
both for the singlet and the triplet channels 
also at finite $\bfq$ there remain one visible and three dark excitons, as in the limit $\bfq=0$ [see Fig. \ref{fig4}(a) and (c)].
This confirms that by suppressing the interlayer hopping the antisymmetric $|FR^{A}_{-}\rangle$ exciton cannot couple with excitons of different parity and continues to be dark.
All the excitons keep the same parity as at $\bfq=0$.
The two dark $|(FR+CT)^{B}_\pm\rangle$ excitons remain degenerate, since the e-h interaction $\hat V$ has no effect on them. 
They are located at lower energies than the $|(FR+CT)^{A}_\pm\rangle$ excitons as $\hat V$ is repulsive.
In particular,  the dark $|(FR+CT)^{A}_-\rangle$ shows a larger dispersion than the visible $|(FR+CT)^{A}_+\rangle$, 
implying that the effect of $\bar I - \bar J$ is larger than $\bar I + \bar J$.
In general, the energy-level ordering is, from the bottom to the top:  $|(FR+CT)^{B}_\pm\rangle$ (degenerate),  $|(FR+CT)^{A}_+\rangle$ and  
$|(FR+CT)^{A}_-\rangle$.

\begin{figure}
\includegraphics[width=0.8\columnwidth, angle=0]{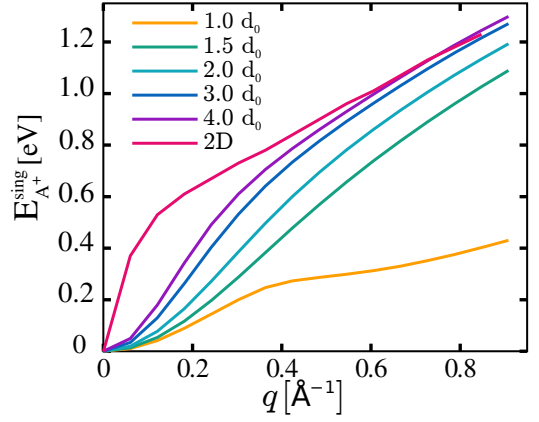}
\begin{center}
\caption{Dispersion of the visible  ``A$^+$'' exciton for different interlayer distances $d$ ($d_0$ is the experimental value).
For each case the exciton energies are defined with respect to the corresponding $q=0$ value.}
\label{fig5}
\end{center}
\end{figure}

By increasing $d$, the screening of  the e-h attraction $W$ is reduced  
and, 
as a consequence, the binding energies of all the excitons increase (however their absolute positions remain almost constant \cite{Wirtz2006}).
In order to directly compare,  for increasing interlayer distances $d$, 
the dispersion of the visible ``A$^+$'' exciton as a function of $q=2\pi/\lambda$, 
in Fig. \ref{fig5} we have hence aligned, for the different separations $d$, 
the exciton energies to their $q=0$ value. 
By increasing the interlayer distance, the dispersion becomes more steep at small $q$ and tends to be the same at large $q$.
As a result of the competition between the e-h exchange interaction and the single-particle band dispersion, 
in the exciton dispersions we can always distinguish two regimes:
(i) at large $q$ (i.e. for $\lambda \ll d$) the sum over $G_z$ in Eq. \eqref{iqa}
can be approximated with an integral. So $\bar{I}^A$ and $\bar{J}^A$  become:
\begin{align}
\bar I^A(\mathbf{q}) \approx & \frac{2\pi}{ q} \beta(q) |S^A(\bfq)|^2
\label{iqa2} \\
\bar J^A(\mathbf{q}) \approx & \bar I^A(\mathbf{q})e^{-qd} \label{jqa2}
\end{align}
with 
\beq 
\beta(q)=\left|\int dz\chi^*_c(z)e^{-qz}\chi_v(z)\right|^2.
\eeq
Under these conditions, as shown in Ref. \onlinecite{Cudazzo2016}, $\bar I^A$ reaches a constant value at large $q$. 
Moreover, since $\lambda \ll d$, the exponential factor in Eq. \eqref{jqa2} goes to zero and $\bar J^A$ becomes negligible. 
As a consequence, in this regime the dispersion of the symmetric and antisymmetric excitons become the same
 and, at large $q$, is  set by the hopping only.
(ii) at small $q$ (i.e. for $\lambda \gg d$) the sums over $G_z$ in Eqs. \eqref{iqa}-\eqref{jqa} is independent of $q$ 
and $S^A(\bfq) = \mathbf{q}\cdot\boldsymbol{\mu}^A$: 
$\bar I$ and $\bar J$ are quadratic in $q$. Therefore in this regime 
the exciton dispersion is also determined by the e-h exchange $\hat {\bar V}$, 
in addition to the hopping contribution that is always present.
At small $q$ the e-h exchange interaction becomes more and more important as $d$ increases, until in the 2D limit
it becomes the dominant contribution 
[compare the dispersion of singlet in Fig. \ref{fig3}(a) and triplet in Fig \ref{fig3}(b)].
Indeed, in the 2D limit, when Eqs. \eqref{iqa2} and \eqref{jqa2} are exact for every $q$,
the e-h exchange contribution becomes linear in $q$, as explained in detail in Ref. \onlinecite{Cudazzo2016}.

\subsection{Plasmon dispersion: long-range Coulomb interaction}

\begin{figure}
\includegraphics[width=0.8\columnwidth, angle=0]{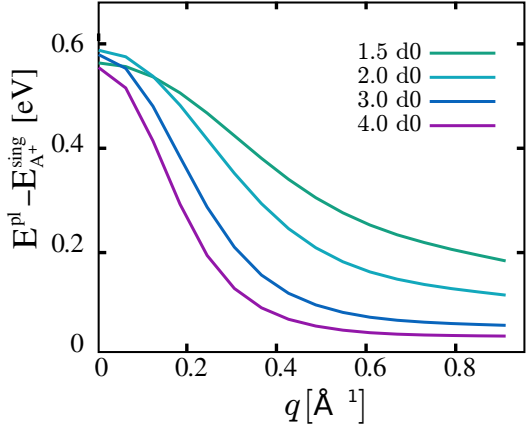}
\begin{center}
\caption{Energy difference between the plasmon and the visible ``A$^+$'' exciton as a function of momentum $\bfq$ for different interlayer distances.}
\label{fig6}
\end{center}
\end{figure}
\begin{figure}
\includegraphics[width=0.8\columnwidth, angle=0]{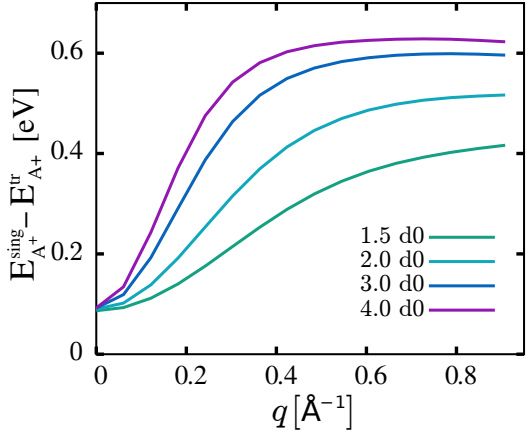}
\begin{center}
\caption{Energy difference between the singlet and triplet ``A$^+$'' excitons as a function of momentum $\bfq$ for different interlayer distances.}
\label{fig7}
\end{center}
\end{figure}

In order to describe the plasmon properties, in the excitonic hamiltonian \eqref{eq1} one has to replace the short-range $\bar v$ with the full Coulomb interaction $v$.
This implies that in the long-range $\mathbf{G}_{||}=0$ contribution to the e-h exchange \eqref{eqv0} also the $G_z=0$ component has to be included.
The excitation transfer interactions [with $\bar I $ and $\bar J$ defined in Eqs. \eqref{iqa}-\eqref{jqa}] thus become:
\begin{align}
 I(\mathbf{q}) =& 
 |S(\bfq)|^2 \frac{4\pi}{q^2} 
 \left| \int dz\chi^*_c(z)\chi_v(z) \right|^2
  + \bar I(\bfq)    \label{iq_pl} \\
 J(\mathbf{q}) =& |S(\bfq)|^2     
 \frac{4\pi}{q^2} \left| \int dz\chi^*_c(z)\chi_v(z) \right|^2 + \bar J(\bfq). \label{jq_pl}
\end{align}

The long-range contribution of the Coulomb interaction 
 is responsible for the difference between the excitation spectra of $\bar L$ and $L$, 
 which are displayed in  Fig. \ref{fig2}(a) and (b), respectively.
 By comparing the poles of $\bar L$ and $L$ we note that the long-range term of $v$ has no effect on the lowest excitons $|(FR+CT)^{B}_\pm\rangle$, since $I=J=0$ for them, 
 and on the antisymmetric exciton $|(FR+CT)^{A}_-\rangle$, as it exactly cancels in the difference $I-J$
[see Eqs. \eqref{iq_pl}-\eqref{jq_pl}].
The repulsive long-range interaction is felt only by the symmetric state $|(FR+CT)^{A}_+\rangle$
 that is the plasmon excitation in $L$. 
As a consequence, its  energy at $\bfq=0$ is upshifted with respect to corresponding ``A$^+$'' pole of $\bar L$ 
by $\sim  (8 \pi/q^2) S(\bfq=0) = 8 \pi |\hat{\mathbf{q}}\cdot\boldsymbol{\mu}|^2 $.
At finite $\bfq$ the plasmon energy displays a quadratic dependence on $\bfq$. 
Without interlayer hopping (i.e. for interlayer spacing $d > 1.5 d_0$), 
the plasmon dispersion is hence similar to that of the triplet exciton energy.
This is a consequence of the cancellation at finite $\bfq$ occurring to a large extent between 
the first and second terms in Eqs. \eqref{iq_pl}-\eqref{jq_pl}.
While the first terms account for the difference between plasmon and singlet exciton (see Fig. \ref{fig6}),
the second terms are responsible for the difference between singlet and triplet excitons (see Fig. \ref{fig7}). 
As a matter of fact, by comparing  Fig. \ref{fig6} and Fig. \ref{fig7} we notice that 
for each interlayer separation
they have an opposite behavior as a function of $q$.

At large  $q$ (i.e. for $\lambda \ll d$) the long-range contribution becomes negligible. 
As shown in Fig. \ref{fig6},
for increasing $d$
the plasmon energy approaches the visible-exciton energy 
for smaller and smaller $q$: the loss function $-\text{Im} \epsilon^{-1}_M$ becomes equal to $\text{Im} \epsilon_M$ when $qd \gg 1$.
In the 2D limit (i.e. $d\rightarrow\infty$), as for any completely isolated system\cite{Sottilephd,Sottile2005}, $-\text{Im} \epsilon^{-1}_M$ and  $\text{Im} \epsilon_M$ 
mathematically coincide for all $\bfq$.

\section{Summary}

From the solution of the {\it ab initio} Bethe-Salpeter equation (BSE) as a function of momentum $\bfq$,
we have obtained the eigenvalue spectrum of the excitonic hamiltonian for the electronic excitations of hexagonal boron nitride
and we have established the connection with measured optical absorption and energy loss spectra. 
We have discussed the properties of both visible and dark excitons on the basis of a simplified model that we have derived from the full {\it ab initio} BSE
and by analogy with the case of molecular solids.
This model has allowed us to provide an efficient description of the excitations in the bulk crystal starting from the knowledge of the excitons in the  single layer.
In this way we have obtained a general picture of the exciton physics in layered materials.
Our analysis uncovers the interplay between the electronic band dispersion and the  electron-hole exchange interaction in setting the exciton properties 
in this important class of materials. Holding a general validity, it can be similarly applied to other van der Waals systems.

\begin{acknowledgments}

This research was supported by the MATRENA Doctoral Programme and Academy of Finland (Contract No. 1260204),
by an {\it \'Energies Durables} Research Grant from the  ́
\'Ecole Polytechnique, the \'Ecole Polytechnique Foundation, and the EDF Foundation,
by a Marie Curie FP7 Integration Grant within the 7th European Union Framework Programme, and
by the European Union's Horizon 2020  research and innovation programme under the Marie Sklodowska-Curie grant agreement No 660695. 
Computational time was granted by GENCI (Project No. 544) and CSC - IT Center for Science.

\end{acknowledgments}

\appendix
\section{Charge-transfer and Frenkel excitons in hBN}
\label{app:a}



In our model we start from the assumption that 
the effective single-particle Hamiltonian that defines the electronic band structure of a layered system can be written as the sum of a single-layer Hamiltonian $\hat{H}_L(\boldsymbol{\rho},z)$ and an effective out-of-plane potential $\delta \hat U(z)$,
which describes the crystal field along the $z$ axis. 
Under these conditions, the band index $n$ and the in-plane wave vector $\mathbf{k}$ that define the eigenstates of $\hat{H}_L(\boldsymbol{\rho},z)$ are also good quantum numbers for the bulk wave function $\psi_{n\mathbf{k},k_z}$  ($k_z$ being the corresponding out-of-plane wave-vector component in the 3D Brillouin zone).
Hence $\psi_{n\mathbf{k},k_z}$ can be expanded in terms of $\phi^i_{n\mathbf{k}}(\boldsymbol{\rho})\chi^i_n(z-\mathbf{R})$ 
($i$ here denotes the layer in the unit cell and $\mathbf{R}$ the lattice vector along $z$). 

For a system characterized by two layers per unit cell, $\phi^i_{n\mathbf{k}}(\boldsymbol{\rho})\chi^i_n(z-\mathbf{R})$ is a set of 2$N$ degenerate states  ($N$ is the number of unit cells) corresponding to the eigenvalues $E_n(\mathbf{k})$ of $\hat{H}_L(\boldsymbol{\rho},z)$ and represent a complete basis set for the bulk wave function. Moreover, in the case of hBN with the AB stacking, the two layers in the unit cell are rotated one with respect
to the other by an angle $\beta=60^{\circ}$ and therefore also the corresponding 2D first Brillouin zones are rotated by the angle $\beta$. 
For a given wave vector $\mathbf{k}$ the in-plane components of the electronic wave functions associated to two inequivalent layers are related by:
\begin{equation}\label{wfp}
\phi^2_{n\mathbf{k}}(\boldsymbol{\rho})=\phi^1_{n\boldsymbol{\beta}\mathbf{k}}(\boldsymbol{\rho})
\end{equation}
where 
$\boldsymbol{\beta}\mathbf{k}$ is the wave vector obtained rotating $\mathbf{k}$ by an angle $\beta$.
Similarly for the corresponding eigenvalues one has: 
\beq \label{eigp}
E^2_n(\mathbf{k})= E^1_n(\beta\mathbf{k}).
\eeq
Choosing the wave vector $\mathbf{k}$ in the first Brillouin zone of the reference layer $i=1$, 
the single-layer basis set is splitted in two subsets of $N$ wave functions $\phi^1_{n\mathbf{k}}(\boldsymbol{\rho})\chi^1_n(z-\mathbf{R})$ with energy $E^1_n(\mathbf{k})$ and $\phi^2_{n\boldsymbol{\beta}^{-1}\mathbf{k}}(\boldsymbol{\rho})\chi^2_n(z-\mathbf{R})$ with energy $E^2_n(\boldsymbol{\beta}^{-1}\mathbf{k})$. 
The ensemble of the two subsets represents a complete basis set for the representation of the bulk wave functions. 
In a more compact notation, the single-layer basis for both excitonic and single-particle Hamiltonians is given by the wave functions $\phi^i_{n\mathbf{k}^{(i)}}(\boldsymbol{\rho})\chi^i_n(z-\mathbf{R})$ with $\mathbf{k}^{(i)}=\mathbf{k}$ for $i=1$ and $\mathbf{k}^{(i)}=\boldsymbol{\beta}^{-1}\mathbf{k}$ for $i=2$.

The single-particle Hamiltonian (written in second quantisation) hence takes the form:
\begin{equation} \label{singleh}
\hat{H}=\sum_{n\mathbf{k}\mathbf{k}'}\sum_{\mathbf{R}\mathbf{S}}\sum_{ij}E^{\mathbf{R}i\mathbf{S}j}_n(\mathbf{k},\mathbf{k}')a^{\dag}_{n\mathbf{k}\mathbf{R}i}a_{n\mathbf{k}'\mathbf{S}j}
\end{equation}
where $E^{\mathbf{R}i\mathbf{S}j}_n(\mathbf{k},\mathbf{k}')$ are the matrix elements of $\hat{H}_L(\boldsymbol{\rho},z)+\delta \hat U(z)$ and are given by the expression:
\begin{equation}
E^{\mathbf{R}i\mathbf{S}j}_n(\mathbf{k},\mathbf{k}')=E^i_n(\mathbf{k})\delta_{\mathbf{R}i,\mathbf{S}j}\delta_{\mathbf{k},\mathbf{k}'}+t^{n\mathbf{k},\mathbf{k}'}_{\mathbf{R}i,\mathbf{S}j}
\end{equation}
with $t^{n\mathbf{k},\mathbf{k}'}_{\mathbf{R}i,\mathbf{S}j}$ denoting the effective interlayer hopping:
\begin{equation}
t^{n\mathbf{k},\mathbf{k}'}_{\mathbf{R}i,\mathbf{S}j}=\int d\boldsymbol{\rho}\phi^{i*}_{n,\mathbf{k}}(\boldsymbol{\rho})\phi^{j}_{n,\mathbf{k}'}(\boldsymbol{\rho})\int dz\chi^{i*}_n(z-\mathbf{R})\delta U(z)\chi^{j}_n(z-\mathbf{S}).
\end{equation}
Defining 
\beq t^n_{\mathbf{R}i,\mathbf{S}j}=\int dz\chi^{i*}_n(z-\mathbf{R})\delta U(z)\chi^{j}_n(z-\mathbf{S}),
\eeq
we have: $t^{n\mathbf{k},\mathbf{k}'}_{\mathbf{R}i,\mathbf{S}i}=t^n_{\mathbf{R}i,\mathbf{S}i}\delta_{\mathbf{k},\mathbf{k}'}$ for $i=j$ and $t^{n\mathbf{k},\mathbf{k}'}_{\mathbf{R}i,\mathbf{S}j}=t^n_{\mathbf{R}i,\mathbf{S}j}\delta_{\boldsymbol{\beta}^{-1}\mathbf{k},\mathbf{k}'}$ for $i\neq j$.
We note that in the present case the hopping $t^{n\mathbf{k},\mathbf{k}'}_{\mathbf{R}i,\mathbf{S}j}$ is not diagonal in $\bfk$ and in this way the
single-particle energies $E^{\mathbf{R}i\mathbf{S}j}_n(\mathbf{k},\mathbf{k}')$ in \eqref{singleh}
  acquire a dependence on both $\bfk$ and $\bfk'$.
    
We consider a two-band system ($n=c,v$) and we take into account only the interlayer hopping between first nearest-neighbour layers ($i\neq j$). In this case  the hopping operators acting on electrons and holes are given by the following expressions:
\begin{align}
\hat{T}_c =& \sum_{\mathbf{R},\mathbf{k}}t^c \left[
a^{\dag}_{c\mathbf{k}\mathbf{R}1}a_{c\boldsymbol{\beta}^{-1}\mathbf{k}\mathbf{R}-\boldsymbol{\tau}2}+a^{\dag}_{c\boldsymbol{\beta}^{-1}\mathbf{k}\mathbf{R}2}a_{c\mathbf{k}\mathbf{R}1} \right] \\
\hat{T}_v =& \sum_{\mathbf{R},\mathbf{k}}t^v \left[
b^{\dag}_{c\mathbf{k}\mathbf{R}1}b_{c\boldsymbol{\beta}^{-1}\mathbf{k}\mathbf{R}-\boldsymbol{\tau}2}+b^{\dag}_{c\boldsymbol{\beta}^{-1}\mathbf{k}\mathbf{R}2}b_{c\mathbf{k}\mathbf{R}1} \right]
\end{align}
where $\boldsymbol{\tau}$ is the smallest lattice vector $(0,0,1)$ and $t^{c(v)}=t^{c(v)}_{\mathbf{R}1,\mathbf{R}-\boldsymbol{\tau}2}=t^{c(v)}_{\mathbf{R}2,\mathbf{R}1}$.
The effect of the hopping is to induce a dispersion along the $z$ axis in reciprocal space and a splitting of the single-layer bands without modifying their in-plane dispersion. 
This is a consequence of the decoupling approximation between in-plane and out-of-plane coordinates.
It is justified by the fact that in hBN the excitons originate from a limited area in the Brillouin zone, so that we can assume that the $k_z$ dispersion in the single-particle band structure is constant for all the relevant $\bfk$ points.

First of all, we neglect the hopping terms in such a way that the charge-transfer and Frenkel excitons are decoupled [see Eq. \eqref{eq1}].
We analyse here the interlayer charge-transfer exciton state, where the electron and the hole are localized on different layers.       
The charge-transfer wave function for the exciton state $\lambda$ is 
\beq
|\Psi^{\lambda}_{ij,\boldsymbol{\tau}}(\mathbf{q})\rangle=\frac{1}{\sqrt{N}}\sum_{\mathbf{R}}|\Psi^{\lambda,ij}_{\mathbf{R},\mathbf{R}+\boldsymbol{\tau}}(\mathbf{q})\rangle
\eeq
with 
\beq 
|\Psi^{\lambda,ij}_{\mathbf{R},\mathbf{R}+\boldsymbol{\tau}}(\mathbf{q})\rangle=\sum_{\mathbf{k}}A^{\lambda,ij,\boldsymbol{\tau}}_{vc,\mathbf{k}}(\mathbf{q})a^{\dag}_{c\mathbf{k}^{(i)}\mathbf{R}i}b^{\dag}_{v\mathbf{k}^{(j)}+\mathbf{q}^{(j)}\mathbf{R}+\boldsymbol{\tau}j}|0\rangle
\eeq
where $\mathbf{k}^{(i)}=\mathbf{k}$ for $i=1$ and $\mathbf{k}^{(i)}=\boldsymbol{\beta}^{-1}\mathbf{k}$ for $i=2$ (the same applies for the wave vector $\mathbf{q}$), while the coefficients $A^{\lambda,ij,\boldsymbol{\tau}}_{vc,\mathbf{k}}(\mathbf{q})$ satisfy the excitonic eigenvalue
equation:
\begin{widetext}
\begin{equation}\label{ct}
[E^i_c(\mathbf{k}^{(i)})-E^j_v(\mathbf{k}^{(j)}+\mathbf{q}^{(j)})]A^{\lambda,ij,\boldsymbol{\tau}}_{vc,\mathbf{k}}(\mathbf{q})-\sum_{\mathbf{k}'}W^{\mathbf{R}+\boldsymbol{\tau}j,\mathbf{R}i}_{\mathbf{R}+\boldsymbol{\tau}j\mathbf{R}i}(v\mathbf{k}^{(j)}+\mathbf{q}^{(j)}c\mathbf{k}^{(i)}v\mathbf{k}^{'(j)}+\mathbf{q}^{(j)}c\mathbf{k}^{'(i)})A^{\lambda,ij,\boldsymbol{\tau}}_{vc,\mathbf{k}'}(\mathbf{q})=E^{\lambda}_{ij,\boldsymbol{\tau}}(\mathbf{q})A^{\lambda,ij,\boldsymbol{\tau}}_{vc,\mathbf{k}}(\mathbf{q})
\end{equation}
\end{widetext}
Here $i$ identifies the index of the layer where the electron of the CT e-h pair is located, while $j$ the layer of the corresponding hole; 
$\boldsymbol{\tau}$ defines the lattice-vector separation along $z$ of the two unit cells to which the layers $i$ and $j$ belong. 
In the following we will focus on the first nearest-neighbour CT states for which $i\neq j$ and $\boldsymbol{\tau}=(0,0,0)$ 
[for the other first nearest-neighbour CT state $\boldsymbol{\tau}$ would be $(0,0,-1)$]. 
In this case we have two possible configurations for the e-h pair: $i$=1 and $j$=2 or $i$=2 and $j$=1. 
They are described respectively by the equations: 
\begin{widetext}
\begin{align}\label{ct12}
[E_c(\mathbf{k})-E_v(\boldsymbol{\beta}^{-1}\mathbf{k}+\boldsymbol{\beta}^{-1}\mathbf{q})]A^{\lambda,12}_{vc,\mathbf{k}}(\mathbf{q})-\sum_{\mathbf{k}'}W(v\boldsymbol{\beta}^{-1}\mathbf{k}+\boldsymbol{\beta}^{-1}\mathbf{q}c\mathbf{k}v\boldsymbol{\beta}^{-1}\mathbf{k}'+\boldsymbol{\beta}^{-1}\mathbf{q}c\mathbf{k}')A^{\lambda,12}_{vc,\mathbf{k}'}(\mathbf{q}) =& E^{\lambda}_{12}(\mathbf{q})A^{\lambda,12}_{vc,\mathbf{k}}(\mathbf{q}) \\
\label{ct21}
[E_c(\boldsymbol{\beta}^{-1}\mathbf{k})-E_v(\mathbf{k}+\mathbf{q})]A^{\lambda,21}_{vc,\mathbf{k}}(\mathbf{q})-\sum_{\mathbf{k}'}W(v\mathbf{k}+\mathbf{q}c\boldsymbol{\beta}^{-1}\mathbf{k}v\mathbf{k}'+\mathbf{q}c\boldsymbol{\beta}^{-1}\mathbf{k}')A^{\lambda,21}_{vc,\mathbf{k}'}(\mathbf{q})=& E^{\lambda}_{21}(\mathbf{q})A^{\lambda,21}_{vc,\mathbf{k}}(\mathbf{q})
\end{align}
where we have dropped the indeces $i$,$j$ since the functional form of both the single-particle energies $E_v$ and $E_c$ and the interlayer effective electron-hole interaction  $W$
is invariant under the exchange of the layer index. 
By applying the rotation $\boldsymbol{\beta}$ to the $\mathbf{k}$ space, Eq. \eqref{ct12} becomes:
\begin{equation}\label{ct12b}
[E_c(\boldsymbol{\beta}\mathbf{k})-E_v(\mathbf{k}+\mathbf{q})]A^{\lambda,12}_{vc,\boldsymbol{\beta}\mathbf{k}}(\boldsymbol{\beta}\mathbf{q})-\sum_{\mathbf{k}'}W(v\mathbf{k}+\mathbf{q}c\boldsymbol{\beta}\mathbf{k}v\mathbf{k}'+\mathbf{q}c\boldsymbol{\beta}\mathbf{k}')A^{\lambda,12}_{vc,\boldsymbol{\beta}\mathbf{k}'}(\boldsymbol{\beta}\mathbf{q})=E^{\lambda}_{12}(\boldsymbol{\beta}\mathbf{q})A^{\lambda,12}_{vc,\boldsymbol{\beta}\mathbf{k}}(\boldsymbol{\beta}\mathbf{q}).
\end{equation}
\end{widetext}
Comparing Eq. \eqref{ct12b} and Eq. \eqref{ct21} we see that, being $E_c(\boldsymbol{\beta}\mathbf{k})=E_c(\mathbf{k})$ (the layer is invariant under rotation of $\pm 60^{\circ}$), the Hamiltonian in Eq. \eqref{ct12b} is the same as in Eq. \eqref{ct21}. This results in the following property for the energies $E^{\lambda}$ and 
coefficients $A^{\lambda}$ of the CT excitonic state:
\begin{eqnarray}
E^{\lambda}_{21}(\mathbf{q}) &=& E^{\lambda}_{12}(\boldsymbol{\beta}\mathbf{q}) \\
A^{\lambda,21}_{vc,\mathbf{k}}(\mathbf{q}) &=& A^{\lambda,12}_{vc,\boldsymbol{\beta}\mathbf{k}}(\boldsymbol{\beta}\mathbf{q}) \label{propct}
\end{eqnarray}

We now analyse the intralayer Frenkel exciton. In this case the excitonic state is  
\beq
|\Psi(\mathbf{q})\rangle=\frac{1}{\sqrt{N}}\sum_{\lambda,\mathbf{R}i}c^{\lambda}_i(\mathbf{q})|\Psi^{\lambda}_{\mathbf{R},i}(\mathbf{q})\rangle,
\eeq
where 
\beq|\Psi^{\lambda}_{\mathbf{R},i}(\mathbf{q})\rangle=\sum_{\mathbf{k}}A^{\lambda,i}_{vc,\mathbf{k}}(\mathbf{q})a^{\dag}_{c\mathbf{k}^{(i)}\mathbf{R}i}b^{\dag}_{v\mathbf{k}^{(i)}+\mathbf{q}^{(i)}\mathbf{R}i}|0\rangle
\eeq 
The electron and the hole of the excitonic pair in this case both belong to the same layer $i$.
The coefficients $A^{\lambda,i}_{vc,\mathbf{k}}(\mathbf{q})$ satisfy the following excitonic eigenvalue equation:
\begin{widetext}
\begin{multline}
[E^i_c(\mathbf{k}^{(i)})-E^i_v(\mathbf{k}^{(i)}+\mathbf{q}^{(i)})]A^{\lambda,i}_{vc,\mathbf{k}}(\mathbf{q})+ \\
\sum_{\mathbf{k}'}\left[\bar{v}^{\mathbf{R}i,\mathbf{R}i}_{\mathbf{R}i\mathbf{R}i}(v\mathbf{k}^{(i)}+\mathbf{q}^{(i)}c\mathbf{k}^{(i)}v\mathbf{k}^{'(i)}+\mathbf{q}^{(i)}c\mathbf{k}^{'(i)})-W^{\mathbf{R}i,\mathbf{R}i}_{\mathbf{R}i\mathbf{R}i}(v\mathbf{k}^{(i)}+\mathbf{q}^{(i)}c\mathbf{k}^{(i)}v\mathbf{k}^{'(i)}+\mathbf{q}^{(i)}c\mathbf{k}^{'(i)}) \right]A^{\lambda,i}_{vc,\mathbf{k}'}(\mathbf{q})=E^{\lambda}_{i}(\mathbf{q})A^{\lambda,i}_{vc,\mathbf{k}}(\mathbf{q}).
\end{multline}
Writing explicitly the eigenvalue equations for the $i=1$ and $i=2$ configurations, we have respectively:
\begin{multline}
\label{eqfr1}
[E_c(\mathbf{k})-E_v(\mathbf{k}+\mathbf{q})]A^{\lambda,1}_{vc,\mathbf{k}}(\mathbf{q})+\sum_{\mathbf{k}'}\left[
\bar{v}
 (v\mathbf{k}+\mathbf{q}c\mathbf{k}v\mathbf{k}'+\mathbf{q}c\mathbf{k}')-W(v\mathbf{k}+\mathbf{q}c\mathbf{k}v\mathbf{k}'+\mathbf{q}c\mathbf{k}')\right]
A^{\lambda,1}_{vc,\mathbf{k}'}(\mathbf{q})=E^{\lambda}_{1}(\mathbf{q}) A^{\lambda,1}_{vc,\mathbf{k}}(\mathbf{q})
\end{multline} 
\begin{multline}\label{eqfr2}
[E_c(\boldsymbol{\beta}^{-1}\mathbf{k})-E_v(\boldsymbol{\beta}^{-1}\mathbf{k}+\boldsymbol{\beta}^{-1}\mathbf{q})]A^{\lambda,2}_{vc,\mathbf{k}}(\mathbf{q})+ \sum_{\mathbf{k}'}\left[ \bar{v}(v\boldsymbol{\beta}^{-1}\mathbf{k}+\boldsymbol{\beta}^{-1}\mathbf{q}c\boldsymbol{\beta}^{-1}\mathbf{k}v\boldsymbol{\beta}^{-1}\mathbf{k}'+\boldsymbol{\beta}^{-1}\mathbf{q}c\boldsymbol{\beta}^{-1}\mathbf{k}') - \right.\\
\left. W(v\boldsymbol{\beta}^{-1}\mathbf{k}+\boldsymbol{\beta}^{-1}\mathbf{q}c\boldsymbol{\beta}^{-1}\mathbf{k}v\boldsymbol{\beta}^{-1}\mathbf{k}'+\boldsymbol{\beta}^{-1}\mathbf{q}c\boldsymbol{\beta}^{-1}\mathbf{k}')\right]
A^{\lambda,2}_{vc,\mathbf{k}'}(\mathbf{q})=
E^{\lambda}_{2}(\mathbf{q})A^{\lambda,2}_{vc,\mathbf{k}}(\mathbf{q})
\end{multline}
Following what we have done for the CT exciton, we apply a rotation $\boldsymbol{\beta}$ to the whole $\mathbf{k}$ space in Eq. \eqref{eqfr2}:
\begin{equation}\label{eqfr2b}
[E_c(\mathbf{k})-E_v(\mathbf{k}+\mathbf{q})]A^{\lambda,2}_{vc,\boldsymbol{\beta}\mathbf{k}}(\boldsymbol{\beta}\mathbf{q})+\sum_{\mathbf{k}'}(\bar{v}(v\mathbf{k}+\mathbf{q}c\mathbf{k}v\mathbf{k}'+\mathbf{q}c\mathbf{k}')-W(v\mathbf{k}+\mathbf{q}c\mathbf{k}v\mathbf{k}'+\mathbf{q}c\mathbf{k}')A^{\lambda,2}_{vc,\boldsymbol{\beta}\mathbf{k}'}(\boldsymbol{\beta}\mathbf{q})=E^{\lambda}_{2}(\mathbf{q})A^{\lambda,2}_{vc,\boldsymbol{\beta}\mathbf{k}}(\boldsymbol{\beta}\mathbf{q}).
\end{equation} 
\end{widetext}
Comparing Eq. \eqref{eqfr2b} with Eq. \eqref{eqfr1} we find that in analogy with the CT state, for the FR exciton the following properties hold: 
\begin{eqnarray}
E^{\lambda}_2(\boldsymbol{\beta}\mathbf{q}) &=& E^{\lambda}_1(\mathbf{q}) \\
A^{\lambda,2}_{vc,\boldsymbol{\beta}\mathbf{k}}(\boldsymbol{\beta}\mathbf{q}) &=& A^{\lambda,1}_{vc,\mathbf{k}}(\mathbf{q}).\label{propfr}
\end{eqnarray}


Finally, we discuss the effect of the hopping, which enters the excitonic hamiltonian through the operator $\hat{T}=\hat{T}_c-\hat{T}_v$, coupling intralayer Frenkel and interlayer charge-transfer states [see Eq. \eqref{eq1}]. 
In particular this coupling is given by the matrix elements of the hopping operator $\hat T$ between $|CT^{\lambda}\rangle$ and $|FR^{\lambda}\rangle$ states. We have the following possibilities:
\begin{widetext}
\begin{eqnarray}
\langle CT^{\lambda}_+|\hat{T}|FR^{\lambda}_+\rangle &=&  \frac{t^c}{2}\left[\sum_{\mathbf{k}}A^{\lambda,1}_{vc,\mathbf{k}}(\mathbf{q})A^{\lambda,21*}_{vc,\mathbf{k}}(\mathbf{q})+\sum_{\mathbf{k}}A^{\lambda,2}_{vc,\mathbf{k}}(\mathbf{q})A^{\lambda,12*}_{vc,\mathbf{k}}(\mathbf{q})\right] \nonumber \\
&-& \frac{t^v}{2}\left[\sum_{\mathbf{k}}A^{\lambda,1}_{vc,\mathbf{k}}(\mathbf{q})A^{\lambda,12*}_{vc,\mathbf{k}}(\mathbf{q})+\sum_{\mathbf{k}}A^{\lambda,2}_{vc,\mathbf{k}}(\mathbf{q})A^{\lambda,21*}_{vc,\mathbf{k}}(\mathbf{q})\right]
\end{eqnarray}
\begin{eqnarray}
\langle CT^{\lambda}_-|\hat{T}|FR^{\lambda}_+\rangle &=&  \frac{t^c}{2}\left[\sum_{\mathbf{k}}A^{\lambda,1}_{vc,\mathbf{k}}(\mathbf{q})A^{\lambda,21*}_{vc,\mathbf{k}}(\mathbf{q})-\sum_{\mathbf{k}}A^{\lambda,2}_{vc,\mathbf{k}}(\mathbf{q})A^{\lambda,12*}_{vc,\mathbf{k}}(\mathbf{q})\right] \nonumber \\
&+& \frac{t^v}{2}\left[\sum_{\mathbf{k}}A^{\lambda,1}_{vc,\mathbf{k}}(\mathbf{q})A^{\lambda,12*}_{vc,\mathbf{k}}(\mathbf{q})-\sum_{\mathbf{k}}A^{\lambda,2}_{vc,\mathbf{k}}(\mathbf{q})A^{\lambda,21*}_{vc,\mathbf{k}}(\mathbf{q})\right]
\end{eqnarray}
\begin{eqnarray}
\langle CT^{\lambda}_+|\hat{T}|FR^{\lambda}_-\rangle &=&  \frac{t^c}{2}\left[\sum_{\mathbf{k}}A^{\lambda,1}_{vc,\mathbf{k}}(\mathbf{q})A^{\lambda,21*}_{vc,\mathbf{k}}(\mathbf{q})-\sum_{\mathbf{k}}A^{\lambda,2}_{vc,\mathbf{k}}(\mathbf{q})A^{\lambda,12*}_{vc,\mathbf{k}}(\mathbf{q})\right] \nonumber \\
&-& \frac{t^v}{2}\left[\sum_{\mathbf{k}}A^{\lambda,1}_{vc,\mathbf{k}}(\mathbf{q})A^{\lambda,12*}_{vc,\mathbf{k}}(\mathbf{q})-\sum_{\mathbf{k}}A^{\lambda,2}_{vc,\mathbf{k}}(\mathbf{q})A^{\lambda,21*}_{vc,\mathbf{k}}(\mathbf{q})\right]
\end{eqnarray}
\begin{eqnarray}
\langle CT^{\lambda}_-|\hat{T}|FR^{\lambda}_-\rangle &=&  \frac{t^c}{2}\left[\sum_{\mathbf{k}}A^{\lambda,1}_{vc,\mathbf{k}}(\mathbf{q})A^{\lambda,21*}_{vc,\mathbf{k}}(\mathbf{q})+\sum_{\mathbf{k}}A^{\lambda,2}_{vc,\mathbf{k}}(\mathbf{q})A^{\lambda,12*}_{vc,\mathbf{k}}(\mathbf{q})\right]\nonumber \\
&+& \frac{t^v}{2}\left[\sum_{\mathbf{k}}A^{\lambda,1}_{vc,\mathbf{k}}(\mathbf{q})A^{\lambda,12*}_{vc,\mathbf{k}}(\mathbf{q})+\sum_{\mathbf{k}}A^{\lambda,2}_{vc,\mathbf{k}}(\mathbf{q})A^{\lambda,21*}_{vc,\mathbf{k}}(\mathbf{q})\right]
\end{eqnarray}
Using the properties of the excitonic coefficients $A^\lambda$ from Eq. \eqref{propct} for the CT state and from Eq. \eqref{propfr} for the FR state, the previous relations become:
\begin{eqnarray}
\langle CT^{\lambda}_+|\hat{T}|FR^{\lambda}_+\rangle &=&  \frac{t^c}{2}\left[\sum_{\mathbf{k}}A^{\lambda,1}_{vc,\mathbf{k}}(\mathbf{q})A^{\lambda,21*}_{vc,\mathbf{k}}(\mathbf{q})+\sum_{\mathbf{k}}A^{\lambda,1}_{vc,\boldsymbol(\beta)\mathbf{k}}(\boldsymbol{\beta}\mathbf{q})A^{\lambda,21*}_{vc,\boldsymbol{\beta}\mathbf{k}}(\boldsymbol{\beta}\mathbf{q}) \right]\nonumber \\
&-& \frac{t^v}{2}\left[\sum_{\mathbf{k}}A^{\lambda,1}_{vc,\boldsymbol{\beta}\mathbf{k}}(\boldsymbol{\beta}\mathbf{q})A^{\lambda,12*}_{vc,\boldsymbol{\beta}\mathbf{k}}(\boldsymbol{\beta}\mathbf{q})+\sum_{\mathbf{k}}A^{\lambda,1}_{vc,\mathbf{k}}(\mathbf{q})A^{\lambda,12*}_{vc,\mathbf{k}}(\mathbf{q})\right]
\end{eqnarray}
\begin{eqnarray}\label{ct-fr+}
\langle CT^{\lambda}_-|\hat{T}|FR^{\lambda}_+\rangle &=&  \frac{t^c}{2}\left[\sum_{\mathbf{k}}A^{\lambda,1}_{vc,\mathbf{k}}(\mathbf{q})A^{\lambda,21*}_{vc,\mathbf{k}}(\mathbf{q})-\sum_{\mathbf{k}}A^{\lambda,1}_{vc,\boldsymbol{\beta}\mathbf{k}}(\boldsymbol{\beta}\mathbf{q})A^{\lambda,21*}_{vc,\boldsymbol{\beta}\mathbf{k}}(\boldsymbol{\beta}\mathbf{q})\right] \nonumber \\
&+& \frac{t^v}{2}\left[\sum_{\mathbf{k}}A^{\lambda,1}_{vc,\mathbf{k}}(\mathbf{q})A^{\lambda,12*}_{vc,\mathbf{k}}(\mathbf{q})-\sum_{\mathbf{k}}A^{\lambda,1}_{vc,\boldsymbol{\beta}\mathbf{k}}(\boldsymbol{\beta}\mathbf{q})A^{\lambda,12*}_{vc,\boldsymbol{\beta}\mathbf{k}}(\boldsymbol{\beta}\mathbf{q})\right]
\end{eqnarray}
\begin{eqnarray}\label{ct+fr-}
\langle CT^{\lambda}_+|\hat{T}|FR^{\lambda}_-\rangle &=&  \frac{t^c}{2}\left[\sum_{\mathbf{k}}A^{\lambda,1}_{vc,\mathbf{k}}(\mathbf{q})A^{\lambda,21*}_{vc,\mathbf{k}}(\mathbf{q})-\sum_{\mathbf{k}}A^{\lambda,1}_{vc,\boldsymbol{\beta}\mathbf{k}}(\boldsymbol{\beta}\mathbf{q})A^{\lambda,21*}_{vc,\boldsymbol{\beta}\mathbf{k}}(\boldsymbol{\beta}\mathbf{q})\right] \nonumber \\
&-&
\frac{t^v}{2}\left[\sum_{\mathbf{k}}A^{\lambda,1}_{vc,\mathbf{k}}(\mathbf{q})A^{\lambda,12*}_{vc,\mathbf{k}}(\mathbf{q})-\sum_{\mathbf{k}}A^{\lambda,1}_{vc,\boldsymbol{\beta}\mathbf{k}}(\boldsymbol{\beta}\mathbf{q})A^{\lambda,12*}_{vc,\boldsymbol{\beta}\mathbf{k}}(\boldsymbol{\beta}\mathbf{q})\right]
\end{eqnarray}
\begin{eqnarray}
\langle CT^{\lambda}_-|\hat{T}|FR^{\lambda}_-\rangle &=&  \frac{t^c}{2}\left[\sum_{\mathbf{k}}A^{\lambda,1}_{vc,\mathbf{k}}(\mathbf{q})A^{\lambda,21*}_{vc,\mathbf{k}}(\mathbf{q})+\sum_{\mathbf{k}}A^{\lambda,1}_{vc,\boldsymbol{\beta}\mathbf{k}}(\boldsymbol{\beta}\mathbf{q})A^{\lambda,21*}_{vc,\boldsymbol{\beta}\mathbf{k}}(\boldsymbol{\beta}\mathbf{q})\right] \nonumber \\
&+& \frac{t^v}{2}\left[\sum_{\mathbf{k}}A^{\lambda,1}_{vc,\mathbf{k}}(\mathbf{q})A^{\lambda,12*}_{vc,\mathbf{k}}(\mathbf{q})+\sum_{\mathbf{k}}A^{\lambda,1}_{vc,\boldsymbol{\beta}\mathbf{k}}(\boldsymbol{\beta}\mathbf{q})A^{\lambda,12*}_{vc,\boldsymbol{\beta}\mathbf{k}}(\boldsymbol{\beta}\mathbf{q})\right].
\end{eqnarray}
\end{widetext}
We can thus conclude that at $\mathbf{q}=0$ the first and second term for each row of Eq. \eqref{ct-fr+} and Eq. \eqref{ct+fr-} cancel each other so that  
$\langle CT^{\lambda}_-|\hat{T}|FR^{\lambda}_+\rangle=\langle CT^{\lambda}_+|\hat{T}|FR^{\lambda}_-\rangle=0$. This means that at $\mathbf{q}=0$ the hopping couples only Frenkel and charge-transfer states of the same parity. As a consequence, also in presence of the hopping the parity of the excitonic states remains a good quantum number. 
Instead, at finite $\mathbf{q}$ there is no more exact cancellation and a mixing between symmetric and antisymmetric states occurs. The parity is no more a good quantum number.


\section{Single-particle band structure}
\label{app:c}

\begin{figure*}
\includegraphics[width=0.8\textwidth, angle=0]{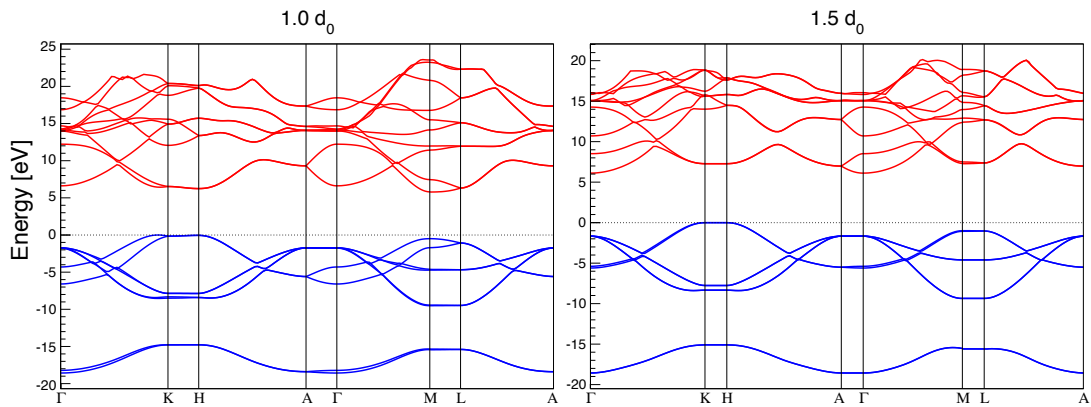}
\begin{center}
\caption{GW electronic band structure for hBN with 1.0 d$_0$ and for 1.5 d$_0$}
\label{fig8}
\end{center}
\end{figure*}

Fig. \ref{fig8} shows the single-particle 
band structures calculated within the GWA for bulk hBN (interlayer distance $d_0$) and for increased interlayer distance $d=1.5 d_0$.

%

\end{document}